\documentclass[a4paper,11pt]{article}
\usepackage[affil-it]{authblk}
\usepackage[english]{babel}
\usepackage{blindtext}
\usepackage{abstract}
\usepackage[dvipdfmx]{graphicx, color}
\usepackage{mathptmx}      % use Times fonts if available on your TeX system
\usepackage{amsmath, amsthm, amssymb}
\usepackage{enumerate}
\usepackage{multirow}
\usepackage{url}
\usepackage{tikz}

\newtheorem{thm}{Theorem}
\newtheorem{prop}{Proposition}

\theoremstyle{definition}
\newtheorem{dfn}{Definition}

\newtheorem{exam}{Example}[section]

\theoremstyle{remark}

\newcommand{\IE}{\textit{i.e., }}
\newcommand{\CF}{\textit{cf.\ }}	% Compare to
\newcommand{\EG}{\textit{e.g., }}

\newcommand{\ETAL}{\textit{et.\ al.\ }}

\newcommand{\abs}[1]{ {\left\lvert #1 \right\rvert} }
\newcommand{\tr}[1]{{\hspace{0.5mm}{}^t\hspace{-0.5mm}#1}}

\def\RR{\mathbb R}

\def\ZZ{\mathbb Z}

\begin{document}                  % DO NOT DELETE THIS LINE

%-------------------------------------------------------------------------
% The introductory (header) part of the paper
%-------------------------------------------------------------------------

% The title of the paper. Use \shorttitle to indicate an abbreviated title
% for use in running heads (you will need to uncomment it).

\title{Ideas of lattice-basis reduction theory for error-stable Bravais lattice determination and ab-initio indexing}
%\shorttitle{Short Title}

% Authors' names and addresses. Use \cauthor for the main (contact) author.
% Use \author for all other authors. Use \aff for authors' affiliations.
% Use lower-case letters in square brackets to link authors to their
% affiliations; if there is only one affiliation address, remove the [a].

\author[1]{R. Oishi-Tomiyasu\thanks{tomiyasu@imi.kyushu-u.ac.jp}}

\affil[1]{%
      Institute of Mathematics for Industry (IMI), Kyushu University}

\date{}  

% Use \shortauthor to indicate an abbreviated author list for use in
% running heads (you will need to uncomment it).

%\shortauthor{Soape, Author and Doe}

% Use \vita if required to give biographical details (for authors of
% invited review papers only). Uncomment it.

%\vita{Author's biography}

% Keywords (required for Journal of Synchrotron Radiation only)
% Use the \keyword macro for each word or phrase, e.g. 
%\keyword{Bravais lattice}\keyword{lattice basis}\keyword{reduction}\keyword{algorithm}\keyword{indexing}

%\keyword{keyword}

% PDB and NDB reference codes for structures referenced in the article and
% deposited with the Protein Data Bank and Nucleic Acids Database (Acta
% Crystallographica Section D). Repeat for each separate structure e.g
% \PDBref[dethiobiotin synthetase]{1byi} \NDBref[d(G$_4$CGC$_4$)]{ad0002}

%\PDBref[optional name]{refcode}
%\NDBref[optional name]{refcode}

\maketitle                        % DO NOT DELETE THIS LINE

%\begin{synopsis}
%	Error-stable Bravais lattice determination algorithms for 2D and 3D lattices are presented.
%\end{synopsis}

\begin{abstract}
In ab-initio indexing, for a given diffraction/scattering pattern, the unit-cell parameters and the Miller indices assigned to reflections in the pattern are determined simultaneously. "Ab-initio" means a process performed without any good prior information on the crystal lattice. Newly developed ab-initio indexing software is frequently reported in crystallography. However, it is not widely recognized that use of a Bravais lattice determination method, which is tolerant to experimental errors, can simplify indexing algorithms and increase their success rates.

One of the goals of this article is to collect information on the lattice-basis reduction theory and its applications.
%For this, the case of 2D lattices, which is simpler than the case of 3D lattices, is mainly used as an example.
The main result is Bravais lattice determination algorithm for 2D lattices, along with a mathematical proof that it works even for parameters containing large observational errors. As in our error-stable algorithm for 3D lattices, it uses two lattice-basis reduction methods that seem to be optimal for different symmetries. 

In indexing, a method for error-stable unit-cell identification is also required to exclude duplicate solutions.
We introduce several methods to measure the difference of unit cells known in crystallography and mathematics.
\end{abstract}

%-------------------------------------------------------------------------
% The main body of the paper
%-------------------------------------------------------------------------
% Now enter the text of the document in multiple \section's, \subsection's
% and \subsubsection's as required.

\section{Introduction}

In Bravais lattice determination, 
the Bravais-type and the parameters of the conventional cell are determined from the parameters of a primitive cell.
%unless auto-indexing is executed assuming the Bravais type.
A set of unit-cell parameters $a, b, c, \alpha, \beta, \gamma$ or a metric tensor ($3$-by-$3$ positive definite symmetric matrix) is used to represent a three-dimensional (3D) lattice.

If the metric tensor is Niggli-reduced and has exact values,
it is known that the Bravais-type and conventional cell 
can be determined by 44 lattice characters (Niggli, 1928; Table~3.1.3.1 of International Tables Vol. A, hereafter abbreviated as ITA (Aroyo, 2016))\nocite{Niggli28}\nocite{Aroyo2016}.
However, this process is required to be error-stable in ab-initio indexing, \IE unit-cell determination from diffraction patterns.

In fact, even if unit-cell parameters containing observation errors is reduced, its true (unknown) parameters may not be reduced for the same basis, although it is \textit{nearly reduced} within a margin of the errors. This situation can lead to a failure in the determination, which cannot be avoided by using parameters represented by integer types.

Bravais lattice determination under experimental uncertainties has been studied by Clegg (1981)\nocite{Clegg81}, Le Page (1982)\nocite{LePage82}, Burzlaff \& Zimmermann (1985)\nocite{Burzlaff85}, Zimmermann \& Burzlaff (1985)\nocite{Zimmermann85} and Andrews \& Bernstein (1988)\nocite{Andrews88}.
The case involving tiny errors, such as rounding errors, has also been discussed
by Buerger (1957)\nocite{Buerger57}, Gruber (1973)\nocite{Gruber73}, K\u{r}iv\'{y} \& Gruber (1976)\nocite{Krivy76}, Zuo \textit{et.\ al.\ } (1995)\nocite{Zuo95}, and Grosse-Kunstleve \textit{et.\ al.\ } (2004)\nocite{Kunstleve2004}.

The method of Zimmermann \& Burzlaff adopted the Delaunay reduction \cite{Delaunay33}, because the number of lattice characters for the Delaunay reduction is 30, which is less than the 44 required for the Niggli reduction.
Andrews and Bernstein (1988) proposed an algorithm that combines the use of 44 lattice characters and Gruber's 25 operations \cite{Gruber73}.
The operations are multiplied recursively to generate nearly Buerger-reduced bases.
The method of SELLA \cite{Andrews2023} uses the 30 lattice characters of Burzlaff \& Zimmermann (1985) to generate nearly Selling-reduced bases.

The method of Oishi-Tomiyasu (2012)\nocite{Tomiyasu2012} is unique in the sense that it is rigorously proven that 
it outputs a (short) finite list of 
all the integer matrices that are required to transform the input unit-cell parameters to its 
centered and reduced form,
as long as 
a mild condition on the magnitude of errors holds. The condition will be explicitly given in Section~\ref{theorems} as ${\mathcal A}_{3,1/2}$.

In the presence of observation errors, the difference between nearly symmetric cells and truly symmetric cells is very ambiguous. 
The algorithm has to output all the candidate types, and leave the determination of the most plausible unit cell to post-processing. 
Since there are only 14 for 3D lattices, it is obvious that the code can output the correct Bravais type (even so, the output for a small threshold gives useful information in practice). 
The non-trivial part is that it can always output the correct reduced basis (more precisely, the basis-transform matrix to obtain it), despite the fact that there are infinitely many lattice bases.

A lack of understanding of the reduction theory might be a reason why error-stable Bravais lattice determination methods are not widely used in crystallography. 
In order to fill the gap, 
we introduced that the same is possible for 2D lattices.
%In particular, the correct basis of the conventional cell is always contained in the output,
%as long as the condition ${\mathcal A}_{2,1}$ is true.

The algorithm for 3D lattices has been used in ab-initio indexing analysis 
software \textit{CONOGRAPH} for powder diffraction (Oishi-Tomiyasu (2014)\nocite{Tomiyasu2014}; \url{https://z-code-software.com/}) and electron backscatter diffraction (Oishi-Tomiyasu (2021)\nocite{Tomiyasu2021}; \url{https://ebsd-conograph.osdn.jp/InstructionsEBSDConograph.html}; migration to GitHub in progress).

In Section~B of the supplementary material, the entire algorithm for 3D lattices is presented for the first time.
It was tested with 4738 CIF files (see the acknowledgements) and confirmed to output the Bravais classes in the CIF files as the highest symmetric solution, except in a few rare cases where the lattice symmetry and space group in the CIF file are not consistent.
The C++ and python codes for 2D and 3D lattices are open source: \url{https://github.com/rtomiyasu}.

Another application of the reduction theory is to find nearly equivalent unit cells.
In Section~6.1, a method using nearly reduced bases is presented.
Section~6.2 introduces a metric on the space of symmetric positive-definite matrices that is known in number theory. 
%In particular, the given distance between unit cells satisfies the triangle inequality. 
%In these methods, a number of basis transforms are applied to obtain all the nearly reduced bases.
%As presented in Section 6.1, for 2D lattices, there is a metric that does not require such transforms, once the reduced basis of the metric tensor is obtained.

The outline of each section is as follows;
Section~2 gives an overview of the ab-initio indexing analyses and describes how Bravais lattice determination and unit-cell identification are used therein.
The basic terms related to Bravais lattices and centring types are defined in Section~3.
Section 4 reviews the reduction methods utilized in this paper, along with some related algorithms. 

Two theorems for $2$-by-$2$ metric tensors containing errors in Section \ref{theorems}, 
were proved in the master thesis of the author's student \cite{Togashi2019}.
An error-stable algorithm for 2D lattices is immediately obtained from the theorems  (Table~\ref{table: algorithm}).
Some results and open problems related to the lattice problems are given in Section~7.

\paragraph{Notation}

For a lattice $L \subset \RR^N$, a \textit{basis} of $L$ is linearly independent vectors $l_1, \ldots, l_n$ ($n \le N$) that satisfy $L = \ZZ l_1 + \cdots + \ZZ l_n$. 
In what follows, it is always assumed that every lattice is full-rank, \IE $n = N$, which is also called the \textit{dimension} of $L$.

For a lattice basis $l_1, \ldots, l_n \subset \RR^n$, the following parallelotope is called a \textit{cell} or \textit{unit cell}:
$$
\{ c_1 l_1 + \ldots + c_n l_n : 0 \le c_i < 1 \}.
$$
The \textit{metric tensor} (or \textit{Gram matrix}, quadratic form) of $l_1, \ldots, l_n$ is defined as the symmetric matrix $(l_i \cdot l_j)_{1 \le i, j \le n}$,
where $u \cdot v$ is the inner product of the Euclidean space.

For a crystal structure, its \textit{primitive lattice}
is the lattice consisting of all the translations that preserve the crystal structure.
The \textit{primitive cell} is the cell spanned by a basis of the primitive lattice.
The \textit{conventional cell} is the cell spanned by the reduced basis of the conventional lattice, which is determined by \textit{centring} as a sublattice of the primitive lattice.
%$L_2$ belong to the same lattice system as $L$ (see \ref{section: Definitions of Bravais classes and lattice systems} for the definition).

%A lattice vector $l \in L$ is a \textit{primitive vector} if it is contained in some basis of $L$, 
%or equivalently if there are coprime integers $m_1, \ldots, m_n$ such that 
%$l = m_1 l_1 + \cdots + m_n l_n$.
As explained in Section \ref{Ideas of lattice-basis reduction},
any lattice-basis reduction method has a prescribed set of conditions that defines the set of \textit{reduced} metric tensors.
A lattice basis is \textit{reduced} if the corresponding metric tensor is reduced.
A cell is \textit{reduced} if the corresponding lattice basis is reduced.

The set of all the $n$-by-$n$ symmetric matrices forms a linear space of dimension $n(n+1)/2$, 
which we denote by ${\mathcal S}^n$.
The open cone consisting of all the positive-definite matrices in ${\mathcal S}^n$ is denoted by ${\mathcal S}^n_{\succ 0}$. 
%${\mathcal S}^n_{\succ 0}$ can be regarded as a cone made by all the $n$-by-$n$ metric tensors.

\section{Application to ab-initio indexing}
\label{Application to ab-initio indexing}

This section explains what types of problems are solved in ab-initio indexing \IE unit-cell determination of crystallography.
In indexing, Bravais-lattice determination and unit-cell identification are applied to multiple primitive cells obtained from a set of input reflections, in order to classify unit cells and eliminate duplicate solutions.

\begin{exam}[Powder ab-initio indexing]
	Figure~\ref{fig:peak search result} shows an example of a powder diffraction pattern.
	In powder patterns, the horizontal coordinates $\blacktriangle$ of peaks give the lengths $\abs{ l }^2$ of reciprocal lattice-vectors $l$.
	No information is given as to which $l$ corresponds to each peak.
	\begin{figure}
		%\begin{tabular}{p{75mm}p{90mm}} 
		%\begin{minipage}[c]{0.5\hsize}
		\scalebox{0.58}{\includegraphics{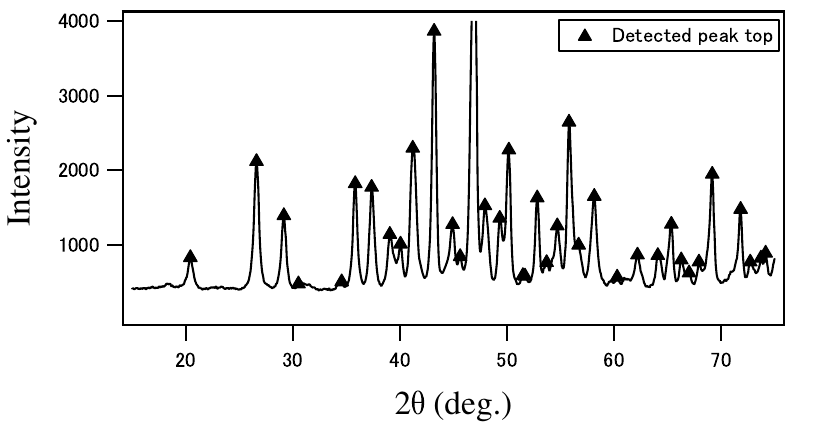}}
		%\end{minipage} &
		%\begin{minipage}[c]{0.5\hsize}
		%\scalebox{0.48}{\includegraphics{fig/PeakSearchResult2.eps}}
		%\end{minipage}
		%\end{tabular}
		\caption{
			Powder diffraction pattern: the triangles $\blacktriangle$ indicate the horizontal coordinates of reflection peaks that provide reciprocal lattice-vector lengths. 
		}
		\label{fig:peak search result}
	\end{figure}
\end{exam}

\begin{exam}[EBSD ab-initio indexing]
	Figure~\ref{fig:peak search result2} shows an example of an EBSD pattern.
	In EBSD patterns, the coordinates of band centerlines provide the orientations $l / \abs{ l }$ of reciprocal lattice-vectors $l$.
	The bandwidths provide the Bragg angles, from which the lengths $\abs{l}$ of the $l$ can be obtained by Bragg's law $n \lambda = 2 d \sin \theta$ and $\abs{l} = 1/d$. The potential errors in bandwidths are much greater than in the centerline coordinates, because of the difficulty in determining (the narrowest) band widths.
	\begin{figure}
		\scalebox{0.6}{\includegraphics{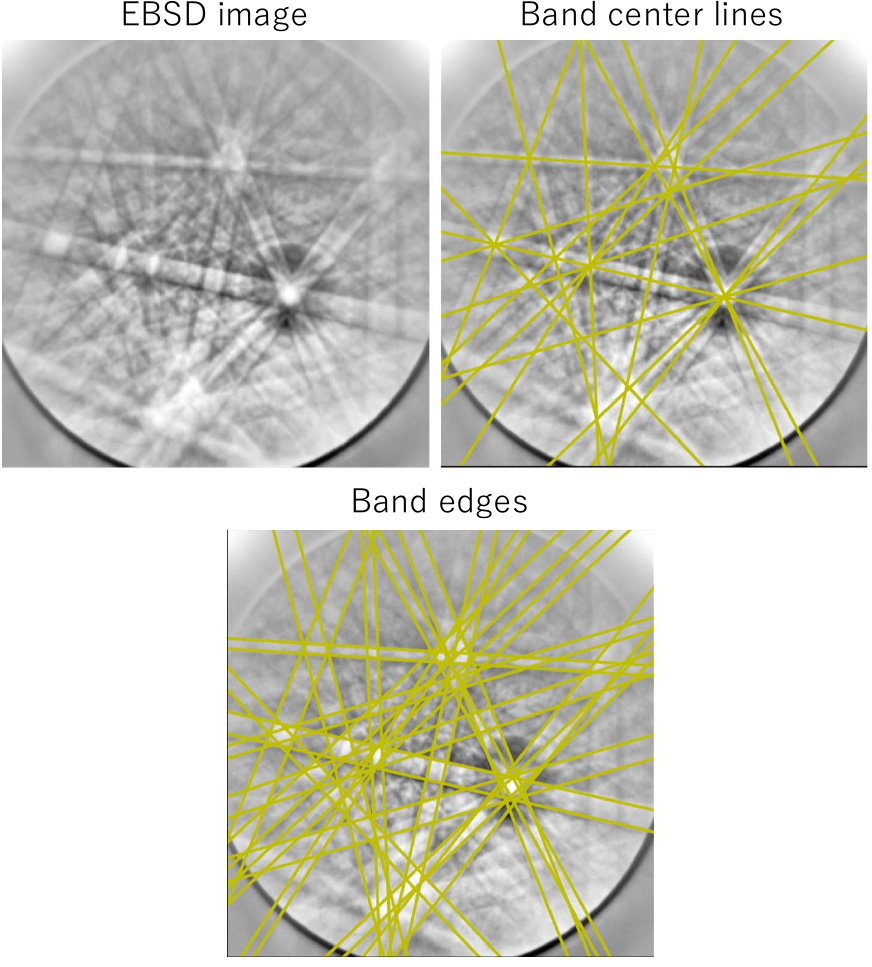}}
		%\scalebox{0.6}{\includegraphics[angle=270]{fig/BandSearchResult.pdf}}
		\caption{
			EBSD pattern: the yellow lines indicate band center lines (upper right) and edges (bottome) that provide reciprocal lattice-vector orientations and lengths.
		}
		\label{fig:peak search result2}
	\end{figure}
\end{exam}

The algorithms of classical ab-initio indexing programs for powder diffraction \textit{ITO} [de Wolff (1958), Visser (1969)], \textit{TREOR} [a trial-and-error method, Kohlbeck \& Horl (1978)] and \textit{DICVOL} [a dichotomy method, Boultif \& Lo{\" u}er (2004)]
\nocite{Wolff58}\nocite{Visser69}\nocite{Kohlbeck78}\nocite{Boultif2004} 
depend on each lattice system and use integer types to avoid the adverse effect of rounding errors.
\textit{TOPAS}\cite{Coelho2003}, \textit{X-cell}\cite{Neumann2003}, \textit{McMaille}\cite{LeBail2004}, and \textit{CONOGRAPH} \cite{Tomiyasu2014} were more recently proposed for powder diffraction.

Ab-initio indexing algorithms and software for EBSD were recently proposed by multiple groups (Li, \ETAL(2014); Oishi-Tomiyasu \ETAL(2021); Nolze \ETAL(2021))\nocite{Li2014}\nocite{Nolze2021}\nocite{Tomiyasu2021}.
The authors of \textit{EBSDL} \cite{Li2015} reported their use of the ITA method, which utilizes lattice characters (\CF Section~\ref{Subspaces of reduced lattices with exact symmetries}).

In ab-initio indexing, a number of nearly identical unit-cell parameters are normally generated from different subsets of reflections, 
which can be used to enhance the success rate of indexing. 
The most plausible unit cell is determined by a figure of merit such as the de Wolff figure of merit \cite{Wolff68} in the process of sorting solutions, which compares observed and calculated reflections.

As a sub-module of indexing software, algorithms for error-stable Bravais-lattice determination and unit-cell identification  
need to be fast enough to process a number of unit cells.
However, these calculations can be time-consuming if nearly reduced bases are searched for, which will be explained in Section~6.1.
For this, our algorithms for error-stable Bravais lattice determination (2D: Section~\ref{theorems}), 3D: section~B) 
avoids directly using nearly reduced bases. 

Nearly identical solutions are normally searched for only in the same Bravais class. 
If unit cells in different Bravais classes are nearly identical, the more symmetric one is usually correct, but it is not appropriate to reject the less symmetric one at this point.

\textit{CONOGRAPH} removes duplicate indexing solutions by a check based on Eq.(\ref{eq: equal or not}) in 
Section~\ref{Removal of duplicate indexing solutions},
while also searching for nearly reduced bases for triclinic cells,
by using nearly Selling-reduced metric tensors (see Definition~\ref{dfn: Selling} for the Selling reduction).
This search is less exhaustive (and thus, less time-consuming) than the methods described in Section~\ref{Removal of duplicate indexing solutions}.

\textit{CONOGRAPH} also uses an idea from the reduction theory in the algorithm for ab-initio indexing \cite{Tomiyasu2013a}.
The algorithm of ITO uses the following identity called Ito's equation (Ito, 1949; de Wolff, 1957; 3.4.3.2 of International Tables Vol. H), which is also known as the parallelogram law\nocite{Ito49}\nocite{Wolff57}\nocite{Gilmore2019}:
\begin{eqnarray}\label{eq: ito equation}
\abs{ l_1 + l_2 }^2 + \abs{ l_1 - l_2 }^2 = 2 (\abs{ l_1 }^2 + \abs{ l_2 }^2).
\end{eqnarray}

In the tiling of ${\mathcal S}^n_{\succ 0}$ provided by Ry{\v s}kov's $C$-type \cite{Ryshkov76}, each facet of the reduced domain can be naturally associated with the set of four vectors $l_1, l_2, l_1 \pm l_2$ in Ito's equation.
Although several reduction methods will be presented in Section~\ref{Ideas of lattice-basis reduction}, many of them have the identical reduced domain (a set of all the reduced metric tensors) as Ry{\v s}kov's $C$-type for 2D and 3D lattices.

\section{Mathematical background: definitions of Bravais classes and lattice systems}
\label{section: Definitions of Bravais classes and lattice systems}

We briefly review the definitions and related terms, primarily for mathematicians who wish to work in crystallography. Most of crystallographers would be at least familiar with the consequences of these theoretical matters. 
For more details, we refer the readers to Michel (1995)\nocite{Michel94}, Zhilinskii (2016)\nocite{Zhilinskii2016} or 1.3.4 of ITA.

For a lattice $L$ with the basis $l_1, \ldots, l_n$,
let $B \in GL_n(\RR)$ be the matrix with column vectors $\begin{pmatrix} l_1 & \ldots & l_n \end{pmatrix}$.
The metric tensor of the lattice basis is the positive-definite symmetric matrix $S := B^T B$.
The automorphism groups of $L$ and $S$ are defined by
\begin{eqnarray*}
	{\rm Aut}(L) &:=& \{ \tau \in O(n) : \tau L = L \}, \\
	{\rm Aut}(S) &:=& \{ \sigma \in GL_n(\ZZ) : \sigma^T S \sigma = S \},
\end{eqnarray*}
where 
$O(n) := \{ U \in GL_n(\RR): U^T U = I \}$ is the orthogonal group.

These automorphic groups are isomorphic, because both are isomorphic to
\begin{eqnarray}\label{eq: automorphism group G}
G := \{ (\tau, \sigma) \in O(n) \times GL_n(\ZZ) : \tau B \sigma^{-1} = B \},
\end{eqnarray}
which can be proved as follows;
the following projections 
\begin{eqnarray*}
	\pi_{1} : (\tau, \sigma) & \mapsto & \tau, \\
	\pi_{2} : (\tau, \sigma) & \mapsto & \sigma.
\end{eqnarray*}
induce group isomorphisms $G \overset{\cong}{\rightarrow} \pi_{1}(G)$ and
$G \overset{\cong}{\rightarrow} \pi_{2}(G)$,
because $\pi_1, \pi_2$ are one-to-one, which can be proved as follows:
\begin{eqnarray*}
	\pi_1(\tau, \sigma) = I
	 & \Leftrightarrow & \tau = I \text{ and } \tau B \sigma^{-1} = B \\
	 & \Leftrightarrow & \tau = \sigma = I, \\
	\pi_2(\tau, \sigma) = I
	 & \Leftrightarrow & \sigma = I \text{ and } \tau B \sigma^{-1} = B \\
	 & \Leftrightarrow & \tau = \sigma = I.
\end{eqnarray*}

We also have
\begin{eqnarray*}
	{\rm Aut}(L) = \pi_{1}(G), \quad
	{\rm Aut}(S) = \pi_{2}(G),
\end{eqnarray*}
because for any $\tau \in O(n)$ and $\sigma \in GL_n(\ZZ)$, 
\begin{eqnarray*}
	\tau \in \pi_{1}(G) &\Leftrightarrow & \tau B \sigma^{-1} = B \text{ for some } \sigma \in GL_n(\ZZ) \\
	& \Leftrightarrow & \tau L = L, \\
	\sigma \in \pi_{2}(G) &\Leftrightarrow& \tau B = B \sigma \text{ for some } \tau \in O(n) \\
	& \Leftrightarrow & \sigma^T B^T B \sigma = B^T B.
\end{eqnarray*}

${\rm Aut}(L)$ ($\cong {\rm Aut}(S)$) can be regarded as a subgroup of the permutation group 
that permutes the lattice-vectors of length less than $M := \max\{ \abs{l_1}, \ldots, \abs{l_n} \}$.
As a result, these automorphism groups are finite groups.

The Bravais class and lattice system of $L$ are defined using these automorphism groups;
for a group $G$,
we shall say that $H_1, H_2 \subset G$ are \textit{conjugate in $G$} if $H_1 = g H_2 g^{-1}$ for some $g \in G$.
From Eq.(\ref{eq: automorphism group G}),
${\rm Aut}(L) = B {\rm Aut}(S) B^{-1}$. Hence, these groups are conjugate in $GL_n(\RR)$.

\begin{dfn}
	Let $L_1$, $L_2 \subset \RR^n$ be lattices with metric tensors $S_1$ and $S_2$. 
	\begin{itemize}
		\item[(1)] $L_1$ and $L_2$ belong to the same \textit{Bravais class (or Bravais type)} 
		if ${\rm Aut}(S_1)$ and ${\rm Aut}(S_2)$ are conjugate in $GL_n(\ZZ)$.
		Equivalently, there is a $g \in GL_n(\ZZ)$ such that the following 
		is a group isomorphism:  
		\begin{eqnarray*}
			{\rm{Aut}}(S_1) & \rightarrow &{\rm{Aut}}(S_{2}), \\
			\sigma & \mapsto & g \sigma g^{-1}.
		\end{eqnarray*}
		
		\item[(2)] $L_1$ and $L_2$ belong to the same \textit{lattice system} 
		if ${\rm Aut}(L_1)$ and ${\rm Aut}(L_2)$ are conjugate in $GL_n(\RR)$.
		Equivalently, there is a $g \in GL_n(\RR)$ such that the following 
		is a group isomorphism:  
		\begin{eqnarray*}
			{\rm{Aut}}(L_1) & \rightarrow &{\rm{Aut}}(L_{2}), \\
			\tau & \mapsto & g \tau g^{-1}.
		\end{eqnarray*}
		
	\end{itemize}
\end{dfn}

Michel (1995) used $O(n)$ in the definition of the lattice systems instead of $GL_n(\RR)$.
This is because 
any subgroups $H_1, H_2 \subset O(n)$ are conjugate in $GL_n(\RR)$ 
if and only if they are in $O(n)$.
In fact, if $H_1 = g H_2 g^{-1}$ for some $g \in GL_n(\RR)$, then $g_2 \in O(n)$ with $H_1 = g_2 H_2 g_2^{-1}$
can be obtained from a singular value decomposition $g = V D \tr{U}$ ($U, V \in O(n)$).

\begin{figure}
	%\begin{tabular}{p{75mm}p{90mm}} 
	%\begin{minipage}[c]{0.5\hsize}
	\scalebox{0.8}{\includegraphics{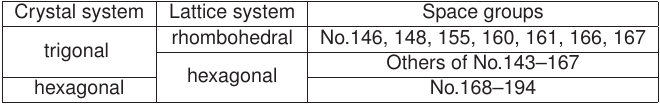}}
	%\end{minipage} &
	%\begin{minipage}[c]{0.5\hsize}
	%\scalebox{0.48}{\includegraphics{fig/PeakSearchResult2.eps}}
	%\end{minipage}
	%\end{tabular}
	\caption{Lattice systems of trigonal/hexagonal crystal systems} 
	\label{table: Lattice systems of trigonal/hexagonal crystal systems} 
\end{figure}

If $H_1, H_2 \subset GL_n(\ZZ)$ are conjugate in 
$GL_n(\ZZ)$, they are also conjugate in $GL_n(\RR)$.
Thus, the classification by Bravais classes gives a refinement of that by lattice systems.
The \textit{crystal system} is defined similarly to the lattice system, by using the symmetry of crystal structures. The lattice and crystal systems agree for $n = 2$ and differ for $n = 3$ only in the part shown in Figure~\ref{table: Lattice systems of trigonal/hexagonal crystal systems}.
%\begin{table}
%\caption{Lattice systems of trigonal/hexagonal crystal systems} 
%\label{table: Lattice systems of trigonal/hexagonal crystal systems} 
%\begin{tabular}{|c|c}
%	Crystal system & Lattice system \\
%	\hline 
%	trigonal & \multirow{2}{*}{rhombohedral} \\ 
%	\cline{1-1} \multirow{2}{*}{hexagonal} &  \\ 
%	\cline{2-2} & hexagonal \\
%\end{tabular}
%\end{table}

For any prime $p \ne 2$, the following map is injective on any 
finite subgroup of $GL_n(\ZZ)$ due to a classical result of Minkowski (1887)\nocite{Minkowski1887}.
\begin{eqnarray*}
	GL_n(\ZZ) & \rightarrow & GL_n(\ZZ / p \ZZ) \\
	Z & \mapsto & Z \mod p.
\end{eqnarray*}

Therefore, the isomorphism classes and the conjugacy classes of finite groups in $GL_n(\ZZ)$
correspond one-to-one to those of the finite group $GL_n(\ZZ/3\ZZ)$,
which implies that the number of Bravais classes is finite for every $n > 0$.

\section{Ideas of lattice basis reduction}
\label{Ideas of lattice-basis reduction}

%The number of basis transforms required for centring is one of the criteria for determining which reduction theory to use.
%Ideally, we want to have a reduction theory that makes the symmetry discernable once the metric tensor is reduced.
%As shown by Oishi-Tomiyasu (2012), the Venkov reductions with respect to $S_0 = A_3$, $I_3$ 
%are as such for face-centered (and body-centered because its reciprocal lattice is face-centered) and primitive monoclinic cells
%but need an additional assumption for rhombohedral and base-centered cells.

In the following, we review some basic facts of the lattice-basis reduction theory.
The metric on ${\mathcal S}^n$ defined in Section~\ref{Metric on the moduli space}
is used to define the Venkov reduction in Section~\ref{Venkov, Selling and Minkowski reductions}.
Minkowski- and Selling reductions of low-rank lattices,
can be regarded as special cases of the Venkov reduction (\CF Propositions~\ref{prop: Minkowski and Venkov}~and~\ref{prop: Selling and Venkov}).

\begin{figure}
	\scalebox{0.6}{\includegraphics{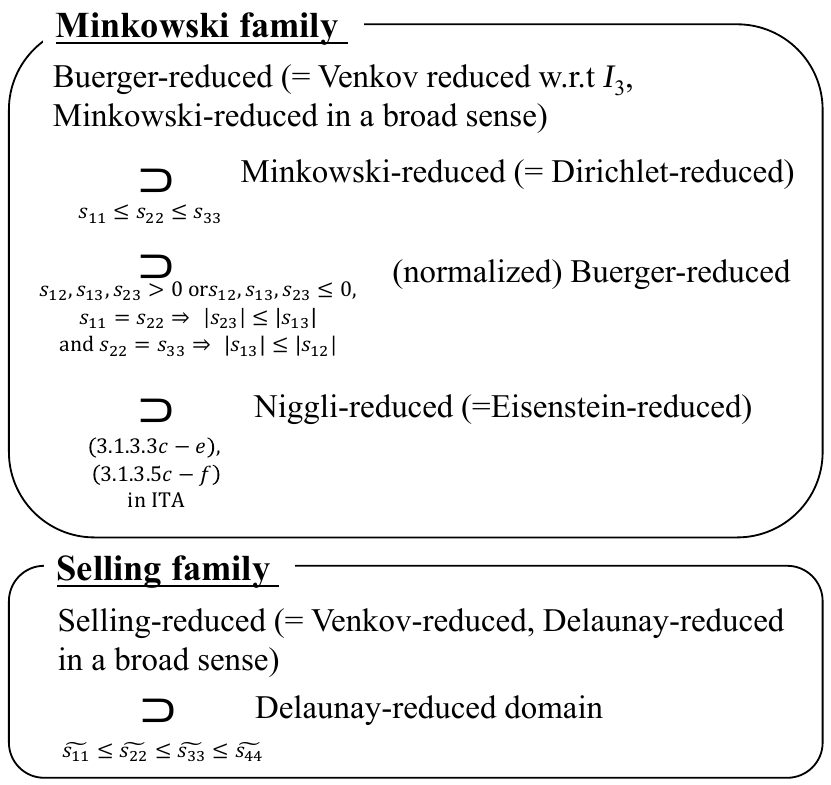}}
	%\scalebox{0.6}{\includegraphics[angle=270]{fig/ReductionMethodsFor3D.pdf}}
	\caption{Inclusion relations of the reduced domains.
	The additional conditions under $\supset$
	are for the reduction in the right-hand side of $\supset$. The $\tilde{s}_{ii}$ are the diagonal entries of Eq.(\ref{eq: tilde{S}}).} 
	\label{fig: Inclusion relations between the reduced domains} 
\end{figure}

Several reduction methods for 3D lattices, described in Section~\ref{Venkov, Selling and Minkowski reductions}, are summarized in Figure~\ref{fig: Inclusion relations between the reduced domains}.

\subsection{Euclidean metric on ${\mathcal S}^n$}
\label{Metric on the moduli space}

%${\mathcal S}^n_{\succ 0}$ can be regarded as the space made by all the $n$-by-$n$ metric tensors.
%As a result, points of the orbit space ${\mathcal S}^n_{\succ 0} / GL_n(\ZZ)$ parametrize the lattices of rank $n$.

On ${\mathcal S}^n$, an inner product is defined for any $S, T \in {\mathcal S}^n$ by 
\begin{eqnarray*}
	S \bullet T &:=& {\rm Trace}(S T)
	= \sum_{i=1}^N \sum_{j=1}^N s_{ij} t_{ij} \\
	&=& \sum_{i=1}^N s_{ii} t_{ii} + 2 \sum_{1 \le i < j \le n}^N \sum_{j=1}^N s_{ii} t_{ii}.
\end{eqnarray*}
Since $S \bullet T$ is a component-wise multiplication, 
$S \bullet T$ is equivalent to the standard inner product of the Euclidean space of dimension $n(n+1)/2$, although the product of non-diagonal entries is multiplied by 2.
%The norm $\abs{ S } := (S \bullet S)^{1/2}$ induced by the inner product 
%s equivalent to the standard Euclidean norm.
In particular, ${\mathcal S}^n$ can be regarded as a Euclidean space with the norm $\abs{ S } := (S \bullet S)^{1/2}$.

From the second equality, $S \bullet T = T \bullet S$ holds. 
The norm $\abs{ S } := (S \bullet S)^{1/2}$ fulfills 
$\abs{ S } \ge 0$, and $\abs{ S } = 0$ if and only if $S = 0$.
The geodesic between $S$ and $T$ is given by the line segment $\lambda S + (1 - \lambda) T$ ($0 \le \lambda \le 1$).
%The distance $\abs{ S - T }$ can be calculated efficiently.
%The metric in Section~\ref{A metric on the modular space of lattices} does not have this drawback, although it is more computationally expensive.

\subsection{Minkowski, Delaunay, Selling, and Venkov reductions}
\label{Venkov, Selling and Minkowski reductions}

For $i = 1, \ldots, n$, let ${\mathbf e}_i = \tr{(0, \ldots, 0, 1, 0, \ldots, 0)}$ be the standard basis of $\RR^n$.
We shall say that $\{ v_1, \ldots, v_i \} \subset \ZZ^n$ is a \textit{primitive set} of $\ZZ^n$ if it is a subset of a basis of $\ZZ^n$.

\begin{dfn}
	$S := (s_{ij})_{1 \leq i, j \leq N} \in {\mathcal S}_{\succ 0}^n$
	is \textit{Minkowski-reduced} if the following equality holds for $i = 1, \ldots, n$:
	\begin{eqnarray}\label{eq:condition1}
		s_{ii} = \min \{ \tr{ v } S v : \{ \mathbf{e}_1, \ldots, \mathbf{e}_{i-1}, v \} \text{ is a primitive set of } \ZZ^n \}.
	\end{eqnarray}
\end{dfn}

The following subset of ${\mathcal S}_{\succ 0}^n$ is the \textit{Minkowski-reduced domain}:
\begin{eqnarray}\label{eq: Minkowski domain}
	{\mathcal D}_{min} := \{ S \in {\mathcal S}_{\succ 0}^n : S \text{ is Minkowski-reduced} \}
\end{eqnarray}

From the definition,
\begin{eqnarray}\label{eq: Dmin}
	{\mathcal D}_{min} = \bigcap_{i = 1}^n \bigcap_{ \{ \mathbf{e}_1, \ldots, \mathbf{e}_{i-1}, v \} \text{: primitive set of } \ZZ^n  } \{ S \in {\mathcal S}_{\succ 0}^n : s_{ii} \le \tr{ v } S v \}.
\end{eqnarray}

The above equality defines ${\mathcal D}_{min}$ as the intersection of infinitely many half spaces,
because $s_{ii} \le \tr{ v } S v$ is a linear inequality on the entries of $S$.
Minkowski (1905)\nocite{Minkowski05} proved that only finitely many of $s_{ii} \le \tr{ v } S v$ are effective in Eq.(\ref{eq: Dmin}). As a result, the topological closure of ${\mathcal D}_{min}$ in ${\mathcal S}^n$ is a polyhedral cone (Theorem 1.3 of Chap.12 of Cassels (1978))\nocite{Cassels78}. 

Regarding to the following, see Lemma 1.2 of Chap.12 in Cassels (1978) for the proof.
\begin{exam}\label{exam: Minkowski reduction}
	For $n \le 4$, $S =(S_{ij}) \in {\mathcal S}^n_{\succ 0}$ is Minkowski-reduced if and only if
	\begin{itemize}
		\item $0 < s_{11} \le s_{22} \le \cdots \le s_{nn}$ and
		
		\item $s_{ii} \le \tr{v} S v$ for any $v = (v_j) \in \ZZ^n$ with $v_j = -1, 0, 1$ for $j < i$, $v_i = 1$ and $v_j = 0$ for $j > i$.
	\end{itemize}
	In particular, $S =(S_{ij}) \in {\mathcal S}^2_{\succ 0}$ is Minkowski-reduced if and only if $0 \leq |2s_{12}| \leq s_{11} \leq s_{22}$.
\end{exam}

For general $n > 0$, 
the following holds for ${\mathcal D} := {\mathcal D}_{min}$ and the group $G := \{ (g_{ij}) \in GL_n(\ZZ) : \text{diagonal and } g_{ii} = \pm 1 \}$ of order $2^n$:
\begin{enumerate}
	\item ${\mathcal S}_{\succ 0}^n = \bigcup_{g \in GL_n(\ZZ)} g {\mathcal D} \tr{g}$.
	
	\item For any $g_1, g_2 \in GL_n(\ZZ)$, $g_1 {\mathcal D} \tr{g}_1 = g_2 {\mathcal D} \tr{g}_2$ if and only if
	$g_1^{-1} g_2 \in G$.
	If $g_1^{-1} g_2 \notin G$, 
	$g_1 {\mathcal D} \tr{g}_1 \cap g_2 {\mathcal D} \tr{g}_2$ is empty, or contained in the boundaries of $g_i {\mathcal D} \tr{g}_i$ ($i = 1, 2$).
	
\end{enumerate}

%By setting $v = {\mathbf e}_{j}$ or $v = {\mathbf e}_{i} + {\mathbf e}_{j}$ ($i < j$),
%it is seen that the entries of $S$ satisfy at least the following:
%\begin{eqnarray}\label{eq:condition2}
%	0 < s_{11} \leq \cdots \leq s_{nn},\ 2 \abs{ s_{ij} } \leq s_{ii}\ (1 \leq i < j \leq n).
%\end{eqnarray}

Example~\ref{exam: Minkowski reduction} imply that for $n \le 4$, if $S \in {\mathcal S}^n_{\succ 0}$ is Minkowski-reduced, then
\[
S \bullet I_{n} \leq (g S \tr{g}) \bullet I_{n} \text{ for any } g \in GL_n(\ZZ),
\]
which will be proved in Proposition~\ref{prop: Minkowski and Venkov}.
The definition of the Venkov reduction is motivated by this.
\begin{dfn}
	$S \in {\mathcal S}^n_{\succ 0}$ is \textit{Venkov-reduced} with respect to a fixed $S_0 \in {\mathcal S}^n_{\succ 0}$ if  
	\[
	S \bullet S_0 \leq (g S \tr{g}) \bullet S_0 \text{ for any } g \in GL_n(\ZZ).
	\]
\end{dfn}

The Venkov-reduced domain is defined by 
\begin{eqnarray}\label{eq: Venkov domain}
	{\mathcal D}_{S_0} := \{ S \in {\mathcal S}_{\succ 0}^n : S \text{ is Venkov-reduced with respect to $S_0$} \}.
\end{eqnarray}

From the equality $(g S \tr{g}) \bullet S_0 = S \bullet (\tr{g} S_0 g)$, we can see that the above 1.\ and 2.\ hold for ${\mathcal D} := {\mathcal D}_{S_0}$ and  
$
G := \{ g \in GL_n(\ZZ) : \tr{g} S_0 g = S_0 \}
$.
${\mathcal D}_{S_0}$ is a polyhedral convex cone for general $n$.
%unlike the case of the Voronoi reduction (Voronoi, 1907; Voronoi, 1908)\nocite{Voronoi07}\nocite{Voronoi08}.

For small dimensions, Minkowski and Venkov reductions can related in the following sense.
\begin{prop}\label{prop: Minkowski and Venkov}
	For $n \le 4$, $S \in {\mathcal S}^n_{\succ 0}$ is Venkov-reduced with respect to $I_n$ if and only if $g S \tr{g}$ is Minkowski-reduced for some permutation matrix $g \in GL_n(\ZZ)$.
\end{prop}
See Section~C of the supplementary material for the proof.
What is called a \textit{Buerger cell} in ITA corresponds to a $3$-by-$3$ Venkov-reduced metric tensor with respect to\ $S_0 = I_3$.
%The \textit{Dirichlet reduction} is the Minkowski reduction for rank 3.

For general dimensions, Selling reduction is defined as a special case of the Venkov reduction (\CF Eq.(11) in \S 5, Chap.2 of Gruber \& Lekkerkerker (1987)\nocite{Gruber87}).
\begin{dfn}\label{dfn: Selling}
	$S \in {\mathcal S}^n_{\succ 0}$ is \textit{Selling-reduced} if $S$ is Venkov-reduced with respect to $A_n = (a_{ij})$ with the entries:
	$$
	a_{ij} =
	\begin{cases}
		2 & \text{if } i = j, \\
		1 & \text{if } i \ne j.
	\end{cases}
	$$ 
	
\end{dfn}

Selling reduction was originally defined for 2D and 3D lattices \cite{Selling1874}
by the following (2) of Proposition~\ref{prop: Selling and Venkov},
which corresponds to the fact that all of the Selling-reduced basis vectors $b_1, \ldots, b_n$ and $b_{n+1} := -b_1- \cdots - b_n$ intersect at obtuse angles.
Such a basis is called \textit{of Voronoi's first kind} in Conway \& Sloane (1992).
The negatives $-\tilde{s}_{ij}$ of the non-diagonal entries of $\tilde{S} = (\tilde{s}_{ij})$ are called \textit{Selling parameters}.
\begin{prop}\label{prop: Selling and Venkov}
	For any $S \in {\mathcal S}^n$, define $\tilde{S} \in {\mathcal S}^{n+1}$ by 
	\begin{equation}\label{eq: tilde{S}}
		\tilde{S} :=
		\begin{pmatrix}
			I_n \\
			\begin{matrix} -1 & \cdots & -1 \end{matrix}
		\end{pmatrix}
		S
		\begin{pmatrix}
			I_n &
			\begin{matrix} -1 \\ \vdots \\ -1 \end{matrix}
		\end{pmatrix}.
	\end{equation}
	
	\begin{enumerate}[(1)]
		\item $S \bullet A_{n} = \tilde{S} \bullet I_{n+1}$ for any $S \in {\mathcal S}^n$.
		
		\item For $n \le 3$, $S \in {\mathcal S}^n_{\succ 0}$ is Selling-reduced if and only if all the non-diagonal entries of $\tilde{S}$ are non-positive.
		
	\end{enumerate}
\end{prop}
(1) can be proved by direct calculation. 
See Section~C for a proof of (2), which uses the reduced domain ${\mathcal D}_{vo}$ provided by the Voronoi theory of perfect forms \cite{Voronoi07}.

\begin{dfn}
	\begin{enumerate}[(i)]
		\item As in Balashov \& Ursell (1957)\nocite{Balashov57}, we shall say that
		$S \in {\mathcal S}^n_{\succ 0}$ is \textit{Delaunay-reduced} if $S$ is Selling-reduced and the diagonal entries of $\tilde{S}$ are in ascending order.
		
		\item As in Oishi-Tomiyasu (2012), we shall say that $S \in {\mathcal S}^n_{\succ 0}$ is \textit{Minkowski-reduced in a broad sense}  if $S$ is Venkov-reduced with respect to~$I_n$ (\CF Proposition~\ref{prop: Minkowski and Venkov}).
		%In addition, if $S$ is Selling-reduced, $S \in {\mathcal S}^n_{\succ 0}$ is \textit{Delaunay-reduced in a broad sense}.
		
		%		In addition, $S$ belonging to ${\mathcal D}_B^+$ or ${\mathcal D}_B^-$ is \textit{Buerger-reduced}.
		%\begin{eqnarray}
		%{\mathcal D}_B^+ &:=& \{ (s_{ij}) \in {\mathcal S}^3_{\succ 0}: 0 < s_{11} \leq s_{22} \leq s_{33}, \nonumber \\
		%& & \hspace{5mm} 0 \leq s_{12}, s_{13} \leq \frac{s_{11}}{2},\ 0 \leq s_{23} \leq \frac{s_{22}}{2} \}, \label{eq:definition of D_B^+} \\
		%{\mathcal D}_B^- &:=& \{ (s_{ij}) \in {\mathcal S}^3_{\succ 0}: 0 < s_{11} \leq s_{22} \leq s_{33}, \nonumber \\
		%& & \hspace{5mm} 0 \leq - s_{12}, - s_{13} \leq \frac{s_{11}}{2},\ 0 \leq -s_{23} \leq \frac{s_{22}}{2}, \nonumber \\
		%& & \hspace{5mm} \abs{ s_{12} + s_{13} + s_{23} } \leq \frac{ s_{11} + s_{22} }{2} \}. \label{eq:definition of D_B^-}
		%\end{eqnarray}
		
	\end{enumerate}
\end{dfn}

The \textit{normalized} Buerger-reduced cells by Gruber (1973) can be obtained 
by imposing the following additional constraints on the Venkov-reduced metric tensors with respect to\ $I_3$:
\begin{itemize}
	\item $s_{11} \le s_{22} \le s_{33}$, 
	\item $s_{12}, s_{13}, s_{23} > 0$ or $s_{12}, s_{13}, s_{23} \le 0$, 
	\item $s_{11} = s_{22} \Longrightarrow \abs{s_{23}} \le \abs{s_{13}}$,
	\item $s_{22} = s_{33} \Longrightarrow \abs{s_{13}} \le \abs{s_{12}}$.
\end{itemize}
In the literature, Buerger-reduced cells refer to the normalized Buerger-reduced cells frequently. We will not use this term to avoid such ambiguity.

\subsection{Algorithms to obtain reduced metric tensors}
\label{Algorithms and code for lattice-basis reduction}

This section summarizes mathematical algorithms for lattice-basis reduction that will be used in the following sections.

The Fincke-Pohst algorithm \cite{Fincke83} takes a metric tensor $S$ and an upper bound $M > 0$ as input and calculates all integer vectors $v$ with 
$\tr{v} S v < M$.
This algorithm 
can be used to calculate various types of reduced bases, the automorphism groups \cite{Plesken97} and the figures of merit in indexing analysis.
It is not an approximation algorithm like the LLL reduction in polynomial time \cite{Lenstra82} and can search for all the shortest vectors.
Although it is practical enough for low-rank lattices (\EG $< 8$),
the computation time increases exponentially with the rank, owing to the intrinsic complexity of the solved problem.

The Gauss algorithm for $2$-by-$2$ metric tensors is presented in Table~\ref{table: Gauss algorithm},
which outputs a Minkowski-reduced metric tensor $g S \tr{g} = (s_{ij})$ with
\begin{eqnarray}\label{eq: Gauss reduction}
	0 \le -2s_{12} \le s_{11} \le s_{22}.
\end{eqnarray}

\begin{table}
	\caption{Gauss algorithm for $2$-by-$2$ metric tensor}
	\label{table: Gauss algorithm}
	\begin{tabular}{lp{75mm}}
		%\begin{tabular}{lp{130mm}}
		\multicolumn{2}{l}{(Input)} \\
		$S = (s_{ij})$: & $2$-by-$2$ metric tensor \\
		\multicolumn{2}{l}{(Output)} \\
		$g \in GL_2(\ZZ)$: & matrix such that $g S \tr{g}$ satisfies Eq.(\ref{eq: Gauss reduction}). \\
		
		\multicolumn{2}{l}{(Algorithm)} \\
		1: & Set $g = I_2$. \\
		2: & Let $m$ be the integer closest to $-s_{12}/s_{22}$.\\
		3: & $S := \begin{pmatrix} 0 & 1 \\ 1 & m \end{pmatrix} S \begin{pmatrix} 0 & 1 \\ 1 & m \end{pmatrix}$, \quad
		$g := \begin{pmatrix} 0 & 1 \\ 1 & m \end{pmatrix} g$. \\
		4: & If $s_{11} > s_{22}$, go to line 2.\\
		5: & If $s_{12} > 0$, multiply the first row of $g$ by $-1$. \\
	\end{tabular}
\end{table}

The Niggli reduction (Eisenstein, 1851; Niggli, 1928)\nocite{Eisenstein1851}\nocite{Niggli28} has been used to represent unit cells uniquely in crystallography.
The algorithm for the Niggli reduction explicitly provided by K{\v r}iv{\' y} \& Gruber (1976)\nocite{Krivy76}
is essentially the same as the Gauss algorithm except for the last steps. 
Balashov \& Ursell (1957)\nocite{Balashov57} provided algorithms to carry out the Delaunay reduction and obtain the Dirichlet-reduced cells from the Delaunay-reduced cells.

The basis-transform matrices used in the Gauss algorithm and by Balashov \& Ursell (1957)
contain parameters determined by solving a minimization problem in each iteration such as
\begin{eqnarray*}
	\begin{pmatrix}
		0 & 1 \\
		1 & m \\
	\end{pmatrix},
	\begin{pmatrix}
		1 & 0 & 0 \\
		m & 1 & 0 \\
		n & 0 & 1 \\
	\end{pmatrix}
\end{eqnarray*}

The reduction algorithms can be simplified by preparing for a finite set containing all the basis-transform matrices to be used, although this increases the number of iterations.
Such an algorithm for Selling reduction is found in Conway (1997)\nocite{Conway97} and also briefly introduced in Andrews \ETAL(2019)\nocite{Andrews2019a}.
The Venkov reduction can be carried out in the same way, and all the basis-transform matrices correspond one-to-one to the facets of ${\mathcal D}_{S_0}$.

\section{Error-stable Bravais lattice determination algorithm for 2D lattices}

This section explains how ideas of the lattice-basis reduction theory is used in the algorithm development.

\subsection{Centring in Bravais lattice determination}
\label{section: Centring type determination in Bravais lattice determination}

%If two lattices $L_1$ and $L_2$ with metric tensors $S_1, S_2$ belong to the same lattice system,
%then their ${\rm Aut}(S_i) \subset GL_n(\ZZ)$ ($i = 1, 2$) are conjugate in $GL_n(\RR)$, and thus, also in $GL_n(\QQ)$.
%This implies that 
%$L_i$ contains a sublattice with the same Bravais class as $L_j$ for $(i, j) = (1, 2)$, $(2, 1)$.
For triclinic symmetry, Niggli reduction is used to fix the basis of a crystal lattice.
For the other symmetries,
lattice-basis reduction is conducted as the pre-processing, and the basis is fixed by \textit{centring}, which can be regarded as a process 
to obtain the reduced basis of the conventional lattice.

\begin{figure}
	\scalebox{0.55}{\includegraphics{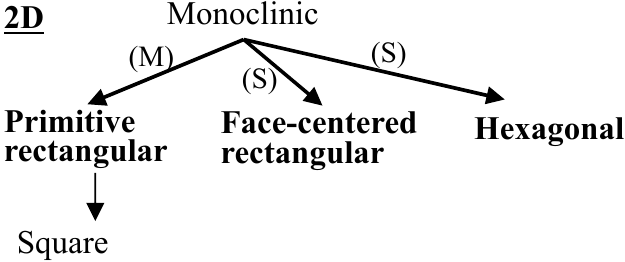}}
	\scalebox{0.49}{\includegraphics{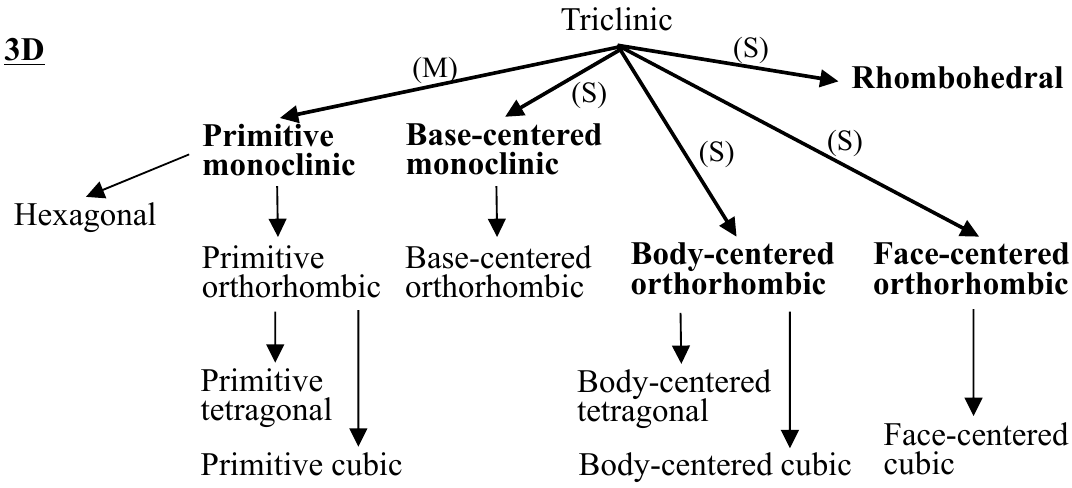}}
	%\scalebox{0.55}{\includegraphics[angle=270]{fig/LatticeSystems2D.pdf}}
	%\scalebox{0.49}{\includegraphics[angle=270]{fig/LatticeSystems3D.pdf}}
	\caption{Lattice system hierarchy used in the algorithm; thick and thin arrows represent the centring and projections to higher symmetric cells, respectively. The basis fixed in centring is essentially unchanged before and after the projections. (M) and (S) refer to the Minkowski and Selling reductions applied before the centring.} 
	\label{table: Lattice systems that correspond to more then one Bravais lattices} 
\end{figure}

Figure~\ref{table: Lattice systems that correspond to more then one Bravais lattices} shows the Bravais-type hierarchy 
we adopted in the algorithms for 2D lattices  (Section~5.4) and 3D lattices (Section~B).
The Bravais lattice determination starts from the centring process.

%, in which 
%basis transforms as follows are applied to an input metric tensor $S$ iteratively until the given conditions are satisfied by $S$.
%$$
%	S \mapsto S := \tr{g} S g.
%$$
%Each lattice-basis reduction method has its own prescribed sets of basis-transforms $g \in GL_n(\ZZ)$,
%and conditions to be satisfied by the reduced metric tensor.
%Since the algorithm terminates in finitely many iterations, any metric tensors have at least one such reduced form.
%Although some may have more, the number of reduced forms does not exceed a certain $M > 0$, which is a specifically determined value for each reduction method.
%In the Niggli reduction, the upper bound $M = 1$, which means that
%the Niggli reduced form is unique for any metric tensors $S$.
%In centring, 
%a faster algorithm can be obtained by choosing reduction methods with $M > 1$,
%since larger $M$ means that we can stop the iterations earlier.
Once the basis is fixed by centring, the remaining process can be conducted just by projecting the parameters to 
a subspace of higher-symmetric metric tensors. 
How close the input metric tensor is to the projection can be measured by their distance.
The norm on ${\mathcal S}^n_{\succ 0}$ 
in Section~\ref{Metric on the moduli space} can be used for this.

\subsection{Subspaces of reduced metric tensors with exact symmetries}
\label{Subspaces of reduced lattices with exact symmetries}

The 44 lattice characters (Niggli, 1928; Table~3.1.3.1 of ITA (Aroyo, 2016))\nocite{Niggli28}
for metric tensors with exact values,
are based on Niggli reduction, 
and consist of the subspaces of metric tensors with exact symmetries and the operations to obtain conventional cells from primitive cells.

\begin{figure}
	\scalebox{0.55}{\includegraphics{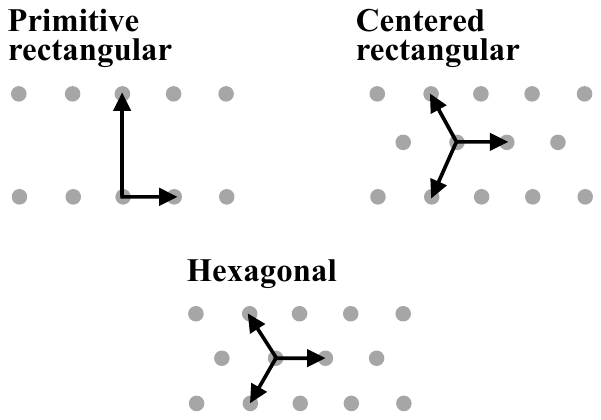}}
	%\scalebox{0.55}{\includegraphics[angle=270]{fig/bases.pdf}}
	\caption{Reduced bases (Minkowski-reduced for the primitive rectangular lattice.  
Any pairs of the three vectors are Selling-reduced for the centered/hexagonal lattice).} 
	\label{fig: Reduced bases} 
\end{figure}

For 2D lattices, our algorithm uses the subspaces given by 
Eqs.(\ref{eq: primitive orthorhombic})--(\ref{eq: hexagonal}),
where $S$ corresponds to a basis taken as in Figure~\ref{fig: Reduced bases}. 
Note that the uniqueness of the reduced basis is up to isometry,
and we can still assume here that the metric tensors have exact values.
\begin{enumerate}[(1)]
	\item Assume that the $2$-by-$2$ metric tensor $S$ is Minkowski-reduced in a broad sense (\IE Venkov-reduced with respect to $I_2$). Then, 
	$S$ is {\bf primitive rectangular} if and only if 
	\begin{eqnarray}\label{eq: primitive orthorhombic}
		S =
		\left(\begin{array}{cc}
			s_{11} & 0 \\
			0 & s_{22}
		\end{array}\right) \text{ for some } s_{11}, s_{22} \in \RR.
	\end{eqnarray}
	
	\item Assume that $S$ is Selling-reduced (\IE Venkov-reduced with respect to\ $A_2$). Then, 
	$S$ is {\bf centered rectangular} if and only if 
	\begin{eqnarray}\label{eq: face-centered orthorhombic}
		S =
		\begin{pmatrix}
			s_{11} & -s_{11}/2 \\
			-s_{11}/2 & s_{22} \\
		\end{pmatrix} 
		\text{ or } 
		\begin{pmatrix}
			s_{22} & -s_{11}/2 \\
			-s_{11}/2 & s_{11} \\
		\end{pmatrix} \\
		\text{ or } 
		\begin{pmatrix}
			s_{22}  &  s_{11}/2 - s_{22}  \\
			s_{11}/2 - s_{22}  & s_{22} \\
		\end{pmatrix} \text{ for some } s_{11}, s_{22} \in \RR. \nonumber  
	\end{eqnarray}
This can be checked by
\begin{small}
	\begin{eqnarray*}
		\begin{pmatrix}
			s_{11} & 0 \\
			0 & 4 s_{22} - s_{11} \\
		\end{pmatrix} 
		=
		\begin{pmatrix}
			1 & 0 \\
			1 & 2 \\
		\end{pmatrix}
		\begin{pmatrix}
			s_{11} & -s_{11}/2 \\
			-s_{11}/2 & s_{22} \\
		\end{pmatrix} 
		\begin{pmatrix}
			1 & 1 \\
			0 & 2 \\
		\end{pmatrix} \\
		=
		\begin{pmatrix}
			0 & 1 \\
			2 & 1 \\
		\end{pmatrix}
		\begin{pmatrix}
			s_{22} & -s_{11}/2 \\
			-s_{11}/2 & s_{11} \\
		\end{pmatrix}
		\begin{pmatrix}
			0 & 2 \\
			1 & 1 \\
		\end{pmatrix} \\
		=
		\begin{pmatrix}
			1 & 1 \\
			1 &-1 \\
		\end{pmatrix}
		\begin{pmatrix}
			s_{22}  &  s_{11}/2 - s_{22}  \\
			s_{11}/2 - s_{22}  & s_{22} \\
		\end{pmatrix}
		\begin{pmatrix}
			1 & 1 \\
			1 &-1 \\
		\end{pmatrix}.
	\end{eqnarray*}
\end{small}

	If $\tilde{S}$ in Eq.(\ref{eq: tilde{S}}) is used,
	Eq.(\ref{eq: face-centered orthorhombic}) is true if and only if
	\begin{eqnarray*}
		\tilde{S} &=&
		g
		\begin{pmatrix}
			s_{11} & -s_{11}/2 & -s_{11}/2 \\
			-s_{11}/2 & s_{22} & s_{11}/2 - s_{22} \\
			-s_{11}/2 & s_{11}/2 - s_{22} & s_{22} \\
		\end{pmatrix} \tr{g} \\
		& & \text{for some permutation matrix in } GL_3(\ZZ).
	\end{eqnarray*}
	Similarly, $S$ is {\bf hexagonal} if and only if
	\begin{eqnarray}\label{eq: hexagonal}
		S =
		\begin{pmatrix}
			s_{11} & -s_{11}/2 \\
			-s_{11}/2 & s_{11} \\
		\end{pmatrix}
		\text{ for some } s_{11} \in \RR,
	\end{eqnarray}
	which is also equivalent to
	\begin{eqnarray*}
		\tilde{S} =
		\begin{pmatrix}
			s_{11} & -s_{11}/2 & -s_{11}/2 \\
			-s_{11}/2 & s_{11} & -s_{11}/2 \\
			-s_{11}/2 & -s_{11}/2 & s_{11} \\
		\end{pmatrix}.
	\end{eqnarray*}
\end{enumerate}

In the algorithm, Minkowski and Selling reductions are both used to account for the efficiency for different symmetries.
For example, $S$ in Eq.(\ref{eq: primitive orthorhombic}) is not close to the boundary of the reduced domains ${\mathcal D}_{I_2}
	=
	\left\{ (s_{ij}) \in {\mathcal S}_{\succ 0}^2: 2 \abs{s_{12}} \le \min\{ s_{11}, s_{22} \}
	\right\}.
$
Unless the error in $s_{12} = 0$ exceeds $s_{11}/2$,
the $S$ remains contained in ${\mathcal D}_{I_2}$.
The same is true for $S$ in Eqs.(\ref{eq: face-centered orthorhombic}) and (\ref{eq: hexagonal}) and ${\mathcal D}_{A_2}$.
The condition ${\mathcal A}_{n,d}$ used in the next section arises from this situation.

\subsection{Theorems for error-stable Bravais-type determination algorithm for 2D lattices}
\label{theorems}

In this section, the observed $2$-by-$2$ metric tensor and its true value 
are denoted by $S^{obs}$ and $S$. As in Proposition~\ref{prop: Selling and Venkov}, $\tilde{S}$ is calculated by Eq.(\ref{eq: tilde{S}}).

The theorems for 2D lattices can be proved under the following condition ${\mathcal A}_{n,d}$ with $n = 2$ and $d = 1$.
The proofs are found in Section~C of the supplementary material.

\paragraph{(${\mathcal A}_{n,d}$)} If the $n$-by-$n$ metric tensor $S$ satisfies $S \bullet T := {\rm Trace}(S T) \geq d (\tr{v} S v)$ for some $0 \ne v \in \ZZ^n$ and $T \in {\mathcal S}^n$, then its observed value $S^{obs}$ also satisfies $S^{obs} \bullet T > 0$.
\vspace{3mm}

In brief, ${\mathcal A}_{2, 1}$ assumes that the upper bound $S \bullet T$ of the squared lattice-vector length $\tr{v} S v$ is never observed as a negative value due to the errors in $S^{obs}$. 
In Oishi-Tomiyasu (2012), the theorems for 3D lattices were proved under ${\mathcal A}_{3,1/2}$. (the condition is the same, although it was stated in a slightly different way).

\begin{thm}[Togashi (2019)] Suppose that $S^{obs}$ is Minkowski-reduced in a broad sense and the Bravais class of $S$ is primitive rectangular. 
	Then, under the assumption ${\mathcal A}_{2,1}$,
	$S$ is also Minkowski-reduced in a broad sense. 
	In particular, $S$ is contained in the linear space:
	$$
	V_{rP} := \left\{\left(\begin{array}{cc}
		s_{11} & 0 \\
		0 & s_{22}
	\end{array}\right) : s_{11}, s_{22}\in \RR \right\}.
	$$
\end{thm}

In the following theorem, 
note that any hexagonal metric tensors are also centered rectangular.

\begin{thm}[Togashi (2019)]\label{thm: face-centered}
	Suppose that $S^{obs}$ is Selling-reduced
	and the Bravais class of $S$ is centered rectangular or hexagonal.
	Under the assumption ${\mathcal A}_{2,1}$,
	$S$ belongs to
	\begin{eqnarray*}
		V_{rC} &:=& \left\{\left(\begin{array}{cc}
			s_{11} & -s_{11}/2 \\
			-s_{11}/2 & s_{22}
		\end{array}\right) : s_{11}, s_{22} \in \RR\right\} \\
		& & \cup\left\{\left(\begin{array}{cc}
			s_{22} & -s_{11}/2 \\
			-s_{11}/2 & s_{11}
		\end{array}\right) : s_{11}, s_{22} \in \RR\right\} \\
		& & \cup\left\{\left(\begin{array}{cc}
			s_{22} & s_{11}/2 - s_{22} \\
			s_{11}/2 - s_{22} & s_{22}
		\end{array}\right) : s_{11}, s_{22} \in \RR\right\}.
	\end{eqnarray*}
	Furthermore, $S$ or $(U g_2) S \tr{(U g_2)}$ is Selling-reduced for the following $U$ and some $g_2 \in GL_2(\ZZ)$ with $\tr{g}_2 A_2 g_2 = A_2$:
	\begin{eqnarray}
		U &:=& \begin{pmatrix} 1 & 0 \\ 0 & - 1 \end{pmatrix}.
		\label{eq: def of U}
	\end{eqnarray}
Note that $g_2 S \tr{g_2}$ is Selling reduced for such a $g_2$ means that $S$ is Selling reduced.
	
	In particular, if $S$ is hexagonal, then $S$ is Selling-reduced and belongs to 
	$$V_{hP} := \left\{\left(\begin{array}{cc}
		s_{11} & -s_{11}/2 \\
		-s_{11}/2 & s_{11}
	\end{array}\right) : s_{11}\in \RR\right\}. 
	$$
	
\end{thm}

The above theorems provide
a (short) finite list of basis-transform matrices required to reduce $S$,  
as long as $S^{obs}$ is reduced and $S$ has the prescribed symmetry.

In the algorithm in Table~\ref{table: algorithm}, the input $S^{obs}$ is reduced in the first step.
Next, $S$ is assumed to have the symmetry of each Bravais class, which is exactly the situation of the theorems.
If $S$ actually has the symmetry,
by applying one of the basis transform matrices $g \in GL_2(\ZZ)$,
$g S \tr{g}$ belongs to the union of the subspaces $V_{rP}$, $V_{rC}$ or $V_{hP}$.
The projection $g S^{obs} \tr{g}$ to each subspace provides an approximation of $g S \tr{g}$.

\begin{table}
	\caption{Error-stable Bravais lattice determination algorithm for 2D lattices
	}
	\label{table: algorithm}
	\begin{tabular}{lp{75mm}}
		%\begin{tabular}{lp{130mm}}
		\multicolumn{2}{l}{(Input)} \\
		$S^{obs}$: & $2$-by-$2$ metric tensor (of a primitive lattice) \\
		\multicolumn{2}{l}{(Output)} \\
		\multicolumn{2}{l}{Ans$_{rP}$, Ans$_{hP}$, Ans$_{rC}$, Ans$_{sP}$:} \\
		 & list of pairs of an integral matrix $g$ and
		the projection of $g S^{obs} \tr{g}$ to a metric tensor with the exact symmetry of the Bravais class. \\
		\multicolumn{2}{l}{(Algorithm)} \\
		(1) & {\bf Gauss reduction}: obtain $g_0 \in GL_2(\ZZ)$ such that $(s_{ij}) := g_0 S^{obs} \tr{g}_0$ satisfies 
		$0 \le -2 s_{12} \le s_{11} \le s_{22}$ by the algorithm in Table~\ref{table: Gauss algorithm}.
		The obtained $(s_{ij})$ is Minkowski- and Selling-reduced. \\ \\
		(2) & {\bf Centring}: prepare the following array for the centered case: \\
		1: &	$C_{rC} :=\left\{ 
		h_F g_0,\
		h_F
		\begin{pmatrix}
			1 & 0 \\
			-1 &-1 \\
		\end{pmatrix} g_0,\
		h_F
		\begin{pmatrix}
			0 & 1 \\
			-1 &-1 \\
		\end{pmatrix}
		g_0
		\right\}^*$, \\
		2: & where $h_F := \begin{pmatrix} 1 & 1 \\ 1 & -1 \end{pmatrix}$ is the matrix to obtain the reduced basis of the centered lattice. \\
		& {\bf (primitive rectangular, rP)} \\
		3: & Compute the $(s_{ij}) := g_0 S^{obs} \tr{g}_0$. \\
		4: & 
		Append 
		$
		\left( g_0,  \begin{pmatrix} s_{11} & 0 \\ 0 & s_{22} \end{pmatrix} \right)
		$
		to Ans$_{rP}$. \\. \\
		& {\bf (hexagonal, hP)} \\ 
		5: & Compute $(s_{ij}) := g_0 S^{obs} \tr{g}_0$ and $s := s_{11} + s_{22} + 2 s_{12}$. \\
		6: & Append $\left( g_0,  \begin{pmatrix} s & -s/2 \\ -s/2 & s \end{pmatrix} \right)$ to Ans$_{hP}$. \\
		& {\bf (centered rectangular, rC)} \\
		7: & for $g$ in $C_{rC}$ do: \\ 
		8: & 
		\hspace{5mm} Compute $(s_{ij}) := g S^{obs} \tr{g}$ and put $S: =
		\begin{pmatrix}
			s_{11} & 0 \\
			0 & s_{22} \\
		\end{pmatrix}$. \\
		9: & \hspace{5mm} Replace $s_{11}$, $s_{22}$ in $S$ and the rows of $g$ if $s_{11} > s_{22}$. \\
		10: & \hspace{5mm} Append $(g, S)$ to Ans$_{rC}$. \\
		11: & end for
		\\ \\
		(3) & {\bf Projection to square (sP)}: \\
		1: & for $(g, S)$ in Ans$_{rP}$ do: \\
		2: & \hspace{5mm} $s := (s_{11} + s_{22})/2$, where $s_{ij}$ is the $(i, j)$-entry of $S$.\\
		3: & \hspace{5mm}
		Append $\left( g, \begin{pmatrix}
			s & 0 \\
			0 & s \\
		\end{pmatrix} \right)$ to Ans$_{sP}$. \\
		4: & end for \\
	\end{tabular}
	\footnotetext[1]{
From Theorem~\ref{thm: face-centered}, $C_{rC}$ is defined as the representatives of the orbit space $G \backslash X$,
where 
$X := \{ h_F h g_2 g_0 : h = I_2 \text{ or } U, \tr{g}_2 A_2 g_2 = A_2 \}$ and
$G$ is the symmetry group of the linear space $\left\{ \begin{pmatrix} s_{11} & 0 \\ 0 & s_{22}  \end{pmatrix} : s_{11}, s_{22} \in \RR \right\}$ generated by $\begin{pmatrix} 0 & 1 \\ 1 & 0 \end{pmatrix}$, $\begin{pmatrix} 1 & 0 \\ 0 & -1 \end{pmatrix}$ and $-I_2$.
	}
\end{table}

The number of candidates output in Ans$_*$
can be reduced by using a suitable distance function $\rm dist$ and threshold $\epsilon$ and
appending only $(g, S^{proj})$ with ${\rm dist}(g S^{obs} \tr{g}, S^{proj}) < \epsilon$ to the list.

At least for the metric on ${\mathcal S}^n_{\succ 0}$ in Section~\ref{Metric on the moduli space},
the projection maps used in (2) and (3) of Table~\ref{table: algorithm} provide the closest point in the subspace to $g S^{obs} \tr{g}$.
Thus, it is concluded that the unknown $S$ fulfills the following inequality for each symmetry $*$: 
$$
	\min_{ (g, S^{proj}) \in \text{Ans}_*} {\rm dist}(g S^{obs} \tr{g}, g S \tr{g}) \ge \min_{ (g, S^{proj}) \in \text{Ans}_*} {\rm dist}(g S^{obs} \tr{g}, S^{proj}).
$$
The value on the right-hand side can be used as a measure of the certainty that $S$ actually has the symmetry $*$.

The projections can be replaced with another one.
For example, if the metric in Section~\ref{A metric on the modular space of lattices} is used, 
the projection to the closest point in $V_{rP}$ is given by
% 1. s11 t^2 = s22 s^2,
% 2. s^2 t^2 = s11 s22 - s12^2.
$$
		\begin{pmatrix}
			s_{11} & s_{12} \\
			s_{12} & s_{22} \\
		\end{pmatrix} \mapsto \sqrt{ s_{11} s_{22} - s_{12}^2 }
		\begin{pmatrix}
			\sqrt{ s_{11} / s_{22} } & 0 \\
			0 & \sqrt{ s_{22}/s_{11} } \\
		\end{pmatrix}.
$$
In this case, the determinant, not the trace, is invariant by the projection.  

If $S^{obs}$ is close to metric tensors in different Bravais classes, 
all of them will be output from the algorithm.
Although the correct Bravais class is usually the most symmetric solution if $\epsilon$ is sufficiently small, 
the selection of the optimal solution should be left to post-processing.
The given algorithm can restrict the candidates for $g$ to a small number of matrices.

The remaining part of this section explains why algorithms for Bravais lattice determination require a condition such as ${\mathcal A}_{2,1}$ and ${\mathcal A}_{3,1/2}$ to ensure that $S^{obs}$ is positive-definite; the reduced domain consisting of all reduced metric tensors (such as ${\mathcal D}_{min}$) contains lower-rank metric tensors in its topological closure as follows:
$$
\begin{pmatrix}
	0 & 0 \\
	0 & 1
\end{pmatrix}, \quad
\begin{pmatrix}
	0 & 0 & 0 \\
	0 & 0 & 0 \\
	0 & 0 & 1 \\
\end{pmatrix}, \quad
\begin{pmatrix}
	0 & 0 & 0 \\
	0 & 1 & 0 \\
	0 & 0 & 1 \\
\end{pmatrix}.
$$
If $S^{obs}$ is nearly equal to a low-rank $S_0$ within a margin of errors, 
$S^{obs}$ is also nearly equal to $g S^{obs} \tr{g}$ for any $g$ with $S_0 = g S_0 \tr{g}$ because we have
$$
S^{obs} \approx S_0 = g S_0 \tr{g} \approx g S^{obs} \tr{g}.
$$
Because $S_0$ is of low rank, infinitely many $g \in GL_n(\ZZ)$ satisfy $S_0 = g S_0 \tr{g}$.
If $S^{obs}$ is reduced as in the theorems, 
then $g S^{obs} \tr{g} \approx S^{obs}$ is also nearly reduced within a margin of errors.
As a result, for such an $S^{obs}$, the algorithm needs to check infinitely many $g \in GL_n(\ZZ)$ as a candidate for which $g S \tr{g}$ might be reduced.
%This is the reason why algorithms for Bravais lattice determination requires an assumption that the errors are not so large that $S^{obs}$ can be nearly low-rank. 

\section{Removal of duplicate indexing solutions}
\label{Use of lattice-basis reduction to measure the difference between two unit-cell parameters}

Sections~\ref{Use of lattice-basis reduction to measure the difference between two unit-cell parameters} and 
\ref{Open problems}
contain some known results in number theory, which are written for expository purposes.

%Section~\ref{Removal of duplicate indexing solutions} outlines how to check whether two unit cells are nearly identical for a given distance function on ${\mathcal S}^n$.
%The distance function in Section~\ref{Metric on the moduli space}
%is primarily used as an example, but
%a metric on ${\mathcal S}^{n} / GL_n(\ZZ)$ introduced in Section~\ref{A metric on the modular space of lattices} is also available.

\subsection{Use of lattice basis reduction to measure the difference between two unit-cell parameters}
\label{Removal of duplicate indexing solutions}

This section starts from a heuristic to compare two unit cells, when their parameters are given.
If the unit cells are in the same Bravais class other than monoclinic and triclinic,
it is straightforward to check whether 
two unit cells are near or not, because their bases are fixed by centring.
Even if the cells are monoclinic or triclinic, if their metric tensors $S = (s_{ij})$ and $T = (t_{ij})$ are both reduced (in the sense of Minkowski, Delaunay, \textit{etc.}), 
it is sufficient in many cases to check the following for a small $\epsilon$:
$$
	\abs{ S - T } \le \epsilon \min\{ \abs{S}, \abs{T} \}
$$
The following component-wise comparison gives a more prudent check to account for the case where the scales of the diagonal entries $s_{ii}$ and $t_{ii}$ largely depend on $i$,
and larger $s_{ii}$ and $t_{ii}$ contain larger errors, which is often the case in crystallography owing to broadening of diffraction peaks
(non-diagonal entries are not directly compared here to account for the case where their values are very close to zero):
\begin{eqnarray}\label{eq: equal or not}
	\abs{ s_{ii} - t_{ii} } & \le &\epsilon \max\{ s_{ii}, t_{ii} \}, \nonumber \\
	& &	\hspace{-22mm} \abs{ (s_{ii} + s_{jj} + 2 s_{ij}) - (t_{ii} + t_{jj} + 2 t_{ij}) } \\
	&\le& \epsilon \max\{ s_{ii} + s_{jj} + 2 s_{ij}, t_{ii} + t_{jj} + 2 t_{ij} \}. \nonumber
\end{eqnarray}

More careful comparison of triclinic and monoclinic cells can be done by checking whether $\abs{ S - g T \tr{g} } \approx 0$ or $\abs{ g S \tr{g} - T } \approx 0$ for some $g \in GL_3(\ZZ)$.
If $S$ and $T$ are both reduced, 
such $g$ provides a nearly reduced basis of $S$ (and $T$).
The operations to generate all the ``nearly reduced'' bases was
finvestigated in Gruber (1973)\nocite{Gruber73}, and utilised in the Bravais-lattice determination method of Andrews \& Bernstein (1988).
If a condition similar to ${\mathcal A}_{n, d}$ is assumed on the error magnitude, 
it is also possible to directly provide all the nearly reduced bases (not a set of generators), as in Oishi-Tomiyasu (2012). The number of potential candidates for nearly reduced bases was 1992 for Niggli and normalized Buerger reductions.

By using essentially the same idea, for an arbitrarily chosen metric ${\rm dist}$ on ${\mathcal S}^n_{\succ 0}$ and a reduced domain ${\mathcal D}_n$, 
a semi-metric (\IE a metric without the triangle inequality) on ${\mathcal S}^n_{\succ 0} / GL_n(\ZZ)$ can be defined by 
\begin{eqnarray}\label{eq: semi-metric}
	\min \left\{ 
	\min\{ {\rm dist}(S, g T \tr{g}), {\rm dist}(g S \tr{g}, T) \} : {\mathcal D}_n \cap g {\mathcal D}_n \tr{g} \ne \emptyset
	\right\}.
\end{eqnarray}

The calculation of the semi-metric Eq.(\ref{eq: semi-metric}) could be much more computationally expensive than Eq.(\ref{eq: equal or not});
for instance, if a metric tensor $S$ is close to $A_3$ (\IE nearly face-centered cubic), 
$g S \tr{g}$ is nearly Buerger-reduced for $336$ distinct $g \in GL_3(\ZZ)$, owing to the fact that  all of $g A_3 \tr{g}$ is Buerger-reduced
(see Appendix~B of Oishi-Tomiyasu (2012) for the calculation).
The number can be mitigated by using the Selling reduction instead of the Buerger reduction, because it has the reduced domain with simpler boundaries.

For 2D lattices, the same calculation is not so computationally expensive,
and can be visualized as follows;
let ${\mathcal D}_{G}$ be the reduced domain of the Gaussian reduction.
\begin{eqnarray*}
	{\mathcal D}_{G} &:=& 
%	\{ S \in {\mathcal S}^2_{\succ 0} : S \text{ is reduced} \} \\
%&=& 
\{ (s_{ij}) \in {\mathcal S}^2_{\succ 0}: 0 \le -2 s_{12} \le s_{11} \leq s_{22} \}.
\end{eqnarray*}

\begin{figure}
	\begin{center}
		\begin{tikzpicture}[scale=1.5, domain=0:3, samples=100, very thick]
			\fill (-0.8,1.7) circle [radius=0.03];
			\fill (-0.2,1.7) circle [radius=0.03];
			\fill (0.2,1.7) circle [radius=0.03];
			\fill (-0.93174, 0.580205) circle [radius=0.03];
			\fill (-0.0682594, 0.580205) circle [radius=0.03];
			\fill ( 0.0682594, 0.580205) circle [radius=0.03];
			\fill (-0.773371, 0.481586) circle [radius=0.03];
			\fill (-0.22663, 0.481586) circle [radius=0.03];
			\draw (0, 0.866)--(0,2.2);
			\draw (-0.5, 0.866)--(-0.5,2.2);
			\draw [domain=-0.5:0] plot(\x,  {sqrt(1-\x*\x)});
			\draw [very thin, dashed] (0.5, 0.866)--(0.5,2.2);
			\draw [very thin, dashed, domain=-0.5:0.5] plot(\x, {sqrt(1-\x*\x)});	
			
			\draw (0,0) node[below left]{0};	%原点
			\draw (-1,0) node[below]{-1};
			\draw (1,0) node[below]{1};
			\draw [thick, ->] (-1.5,0)--(1.2,0) node[right] {$x$};	%x軸
			\draw [thick, ->] (0,-0.5)--(0,2.2) node[above] {$y$};	%y軸
			
			\draw [very thin, dashed] (-1.0, 0.0)--(-1.0,2.2);
			\draw [very thin, dashed] (-0.5, 0.5)--(-0.5,2.2);
			\draw [very thin, dashed] (0.5, 0.887)--(0.5,2.2);
			\draw [very thin, dashed, domain=-1.0:0.0] plot(\x,  {sqrt(1-\x*\x)});
			\draw [very thin, dashed, domain=0.0:0.5] plot(\x,  {sqrt(1-\x*\x)});
			\draw [very thin, dashed, domain=-1.0:0.0] plot(\x,  {sqrt(1-(\x+1)*(\x+1)});
			\draw [very thin, dashed, domain=-1.0:-1.0] plot(\x,  {sqrt(1-(\x+2)*(\x+2)});
			\draw [very thin, dashed, domain=-1.0:0.0] plot(\x,  {sqrt(1/4-(\x+1/2)*(\x+1/2)});
			\draw [very thin, dashed, domain=0.0:0.5] plot(\x,  {sqrt(1-(\x-1)*(\x-1)});
		\end{tikzpicture}
	\end{center}
	\label{fig: upper half plane}
	\caption{
		The area surrounded by solid lines is the reduced domain ${\mathcal D}_{G}$.
		The black dots are points equivalent by the action of $GL_2(\ZZ)$.
	}
\end{figure}
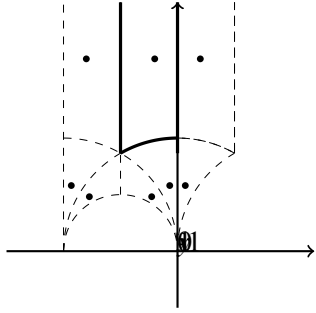

By equating all $c S$ ($0 < c \in \RR$) for each $S$, ${\mathcal D}_{G}$
can be projected onto the upper half plane as in Figure~\ref{fig: upper half plane}.
The following one-to-one correspondence between ${\mathcal S}^2_{\succ 0} / \RR_+$ and the upper half plane ${\mathbb H}$ is given by the projection:
$$
c \begin{pmatrix} 1 & x \\ x & x^2 + y^2 \end{pmatrix} \mapsto x + i y.
$$
The Poincar{\' e} metric $d x d y / y^2$ on ${\mathbb H}$ is known as an invariant metric by the action of $GL_2(\ZZ)$ (the action induced by $S \mapsto g S \tr{g}$ coincides with linear fractional transformations).

As shown in Figure~\ref{fig: upper half plane}, if $g$ and $-g$ are counted as one, 
there are seven $g$ such that
${\mathcal D}_G \cap g {\mathcal D}_G \tr{g} \ne \emptyset$.

However, as pointed out in Kurlin (2022)\nocite{Kurlin2022}, a simpler metric can be defined for 2D lattices, which does not need to account for nearly reduced bases as is done in Eq.(\ref{eq: semi-metric}). 
The metric is obtained just by applying the Gauss algorithm to the metric tensors, and embedding them in $\RR^3$ by
\begin{eqnarray}\label{eq: embedding into R^3}	
	\begin{pmatrix} s_{11} & s_{12} \\ s_{12} & s_{22} \end{pmatrix} \mapsto (s_{11}, s_{22}, s_{11}+2s_{12}+s_{22}).
\end{eqnarray}	

In Eq.(\ref{eq: embedding into R^3}), $s_{11} \le s_{22} \le s_{11}+2s_{12}+s_{22}$ are the values of the \textit{vornorm} map (Conway \& Sloane, 1992; \CF problem 5~of Section \ref{Open problems}).
They characterize 2D lattices uniquely as a continuous invariant, 
because they are equal to the squared lengths of the shortest primitive vectors (Theorem~7 of Conway \& Sloane (1992))\nocite{Conway92}.

As a result, the distance on ${\mathcal S}^2_{\succ 0} / GL_2(\ZZ)$ can be defined using any metrics on $\RR^3$ along with the above map,
which does not require any consideration of nearly reduced bases, once the reduced basis of the metric tensor is obtained.

The group of Kurlin (2022) is also working on its extensions to higher dimensions and periodic point sets (Widdowson \& Kurlin, 2022; Anosova \& Kurlin, 2021)\nocite{NEURIPS2022_9c256fa1}\nocite{10.1007/978-3-030-76657-3_16}.
Similar results on the metrics in the generalized spaces should be useful for the foundation of computational crystallography.

\subsection{A metric on the space of $nD$ lattices}
\label{A metric on the modular space of lattices}

From the norm $\abs{ S } := (S \bullet S)^{1/2}$ defined in Section~\ref{Metric on the moduli space}, a semi-metric (\IE a metric without the triangle inequality) on the orbit space ${\mathcal S}^n_{\succ 0} / GL_n(\ZZ)$
can be defined by 
$$
	d(S, T) = \min_{g_1, g_2 \in GL_n(\ZZ) } \abs{ g_1 S \tr{g}_1 - g_2 T \tr{g}_2 }.
$$

A metric on ${\mathcal S}^n_{\succ 0} / GL_n(\ZZ)$ that satisfies the triangle inequality can be obtained from a metric on ${\mathcal S}^n_{\succ 0}$ with the following invariance property. 
\begin{eqnarray}\label{eq; new metric}
	{\rm dist}_{{\mathcal S}^n_{\succ 0}}(S, T) = {\rm dist}_{{\mathcal S}^n_{\succ 0}}(g S \tr{g}, g T \tr{g}) \text{ for any } S, T \in {\mathcal S}^n.
\end{eqnarray}
It is left to reader as an exercise that the metric on ${\mathcal S}^n_{\succ 0} / GL_n(\ZZ)$ induced by Eq.(\ref{eq: distance}) actually fulfills 
all the axioms of a metric if Eq.(\ref{eq; new metric}) is true.

The metric on ${\mathcal S}^n$ defined below satisfies the above invariance property and provides a typical example of Hadamard spaces, Riemannian symmetric spaces and homogeneous spaces in number theory (Chap.~4.4 of Terras, 1988; chap.~XII of Lang, 1998)\nocite{Terras88}\nocite{Lang98}.
This metric is also known as an \textit{affine-invariant metric} in the field of computer vision \cite{Quang2017}.
A metric on ${\mathcal S}^{n}_{\succ 0} / GL_n(\ZZ)$ induced by Eq.(\ref{eq: distance}), 
can be used to measure the difference between two $n$D lattices specified by their metric tensors. 

The tangent space at each $S \in {\mathcal S}^{n}_{\succ 0}$ can be identified with ${\mathcal S}^{n}$. For any $T_1, T_2 \in {\mathcal S}^{n}$, an inner product and its corresponding norm are defined by
\begin{eqnarray*}
	\langle T_1, T_2 \rangle_{S} &=& {\rm Trace}(S^{-1} T_1 S^{-1} T_2), \\
	\abs{ T_i }_{S} &=& \left( \langle T_i, T_i \rangle_{S} \right)^{1/2}.
\end{eqnarray*}

By the action of $g \in GL_n(\RR)$,
$S$ and $T_i$ ($i = 1, 2$) are mapped to $g S \tr{g}$ and $g T_i \tr{g}$, respectively.
The inner product is invariant by the action, as seen from the following:
\begin{eqnarray}\label{eq: g invariant}
	\langle g T_1 \tr{g}, g T_2 \tr{g} \rangle_{g S \tr{g}} &=& {\rm Trace}(\tr{g}^{-1} S^{-1} T_1 S^{-1} T_2 \tr{g}) \\
	&=& {\rm Trace}(S^{-1} T_1 S^{-1} T_2) = \langle T_1, T_2 \rangle_{S}. \nonumber 
\end{eqnarray}

The length of a curve $\alpha(t): [a, b] \rightarrow {\mathcal S}^n_{\succ 0}$ is given by
$$
\int_{a}^b \abs{ \alpha^\prime(t) }_{\alpha(t)} dt.
$$
The action of any $g \in GL_n(\RR)$ gives an isometry on ${\mathcal S}^n_{\succ 0}$ because 
Eq.(\ref{eq: g invariant}) implies:
$$
\int_{a}^b \abs{ g \alpha^\prime(t) \tr{g} }_{ g \alpha(t) \tr{g} } dt
= \int_{a}^b \abs{ \alpha^\prime(t) }_{\alpha(t)} dt.
$$

In particular, 
for any $L \in GL_n(\RR)$ with $S_1 = L \tr{L}$, 
the length of the geodesic between $S_1$ and $S_2 \in {\mathcal S}^n_{\succ 0}$ is equal to 
that of the geodesic $\ell$ between $I$ and $T := L^{-1} S_2 \tr{L}^{-1}$,
as a result of the invariance property.
The $\ell$ can be parametrized by using 
the eigenvalue decomposition $T= U D \tr{U}$ ($U \in O(n)$) and 
the eigenvalues $d_1, \ldots , d_n > 0$ in the diagonal entries of $D$: 
\begin{eqnarray*}
	\ell(t) &:=& U {\mathcal D}(t) \tr{U} \quad (0 \le t \le 1), \\
	{\mathcal D}(t) &:=&
	\begin{pmatrix} 
		\exp( t \log d_1 ) \\
		& \ddots \\
		& & \exp( t \log d_n )
	\end{pmatrix}.
\end{eqnarray*}
A more detailed discussion can be found in the chapter ``Bruhat-Tits spaces'' of Lang (1999)\nocite{Lang99}.
%The eigenvalues $d_1, \ldots, d_n$ can be obtained as the roots of $\det( S_1 t - S_2 ) = \det S_1 \det (t I - T)$.
As a result, the distance between the above $S_1$ and $S_2$ is equal to
\begin{small}
\begin{eqnarray*}
	{\rm dist}_{{\mathcal S}^n_{\succ 0}}(S_1, S_2)&=&
	\int_{0}^1 \abs{ U \frac{d {\mathcal D}(t)}{dt} \tr{U} }_{ U {\mathcal D}(t) \tr{U}  } dt \nonumber \\
	&=& \int_{0}^1 \left( {\rm Trace} \left( {\mathcal D}(t)^{-1} \frac{d {\mathcal D}(t)}{dt} {\mathcal D}(t)^{-1} \frac{d {\mathcal D}(t)}{dt} \right) \right)^{1/2} d t \nonumber \\
	&=& \left( \sum_{i=1}^n (\log d_i)^2 \right)^{1/2}.
\end{eqnarray*}
\end{small}

Let $[S], [T] \in {\mathcal S}^n_{\succ 0} / GL_n(\ZZ)$ be the orbits (equivalent classes) of $S$ and $T \in {\mathcal S}^n_{\succ 0}$.
The distance on the orbit space is induced by the above distance: 
\begin{eqnarray}\label{eq: distance}
	{\rm dist}_{{\mathcal S}^n_{\succ 0}/GL_n(\ZZ)}([S], [T]) &:=&
	\min_{g \in GL_n(\ZZ)} {\rm dist}_{{\mathcal S}^n_{\succ 0}}(S, g T \tr{g}).
\end{eqnarray}

The minimum value can be calculated as follows (its improvement in the efficacy should be left to a future study. \CF Problem 3 in Section~\ref{Removal of duplicate indexing solutions}); we may assume that $S$ and $T$ are Venkov-reduced with respect to\ $I_n$.
Let $S = L \tr{L}$ be the Cholesky decomposition of $S$.
If some $g \in GL_n(\ZZ)$ satisfies ${\rm dist}_{{\mathcal S}^n_{\succ 0}}(S, g T \tr{g}) < D  := {\rm dist}_{{\mathcal S}^n_{\succ 0}}(S, T)$, 
then all the eigenvalues of $L^{-1} g T \tr{g} \tr{L}^{-1}$ must be less than $e^D$.
If $d_0$ is the minimum eigenvalue of $L^{-1} \tr{L}^{-1}$, then, for any $0 \ne v \in \RR^n$,
\begin{eqnarray*}
e^D & \ge & \frac{ \tr{v} L^{-1} g T \tr{g} \tr{L}^{-1} v }{ \abs{ v }^2 }, \\
d_0 & \le & \frac{ \abs{ \tr{L}^{-1} v }^2 }{ \abs{ v }^2 }.
\end{eqnarray*}
Hence, 
\begin{eqnarray*}
\frac{ e^D }{ d_0 } & \ge & \frac{ \tr{v} g T \tr{g} v }{ \abs{ v }^2 } \ \text{ for any $0 \ne v \in \RR^n$. }
\end{eqnarray*}

Therefore, every row vector $u$ of $g$ must satisfy $u S \tr{u} \le e^D / d_0$.
All such $g \in GL_3(\ZZ)$ can be enumerated by using the Fincke-Pohst algorithm.
The computation time is clearly exponential of rank $n$. 

This affine-invariant metric on ${\mathcal S}^n_{\succ 0}$ is known to be time-consuming in computer vision, although the invariant property is not required in the field.
Therefore, Eq.(\ref{eq: distance}) is practical only for small dimensions.

\section{Open problems}
\label{Open problems}

The following are open problems the author came across in the course of this writing.
1~and~2 relate to algorithms for error-stable Bravais-lattice determination. 
For this, nothing has been done for more than 3 dimensions.

With respect to 2, which reduction method (or $S_0$ for the Venkov reduction) is the most suitable can be estimated from how far the subspace formed by the symmetric metric tensors is from the boundaries of the reduced domain. Our algorithm for 3D lattices of base-centered and rhombohedral centring types might be improved by doing this. 

3 and 4 concern metrics on ${\mathcal S}^n_{\succ 0} / GL_n(\ZZ)$ for the unit-cell identification. Efficient algorithms for this are required in a wide range of problems in crystallography including database searching and indexing.
3~is straightforward for small $n$.
4~is suggested by the metric given by Eq.(\ref{eq: embedding into R^3}) for 2D lattices.
This metric is also reminiscent of the Conway-Sloan conjecture listed as 5. 
\begin{itemize}
	\item[1.] Error-stable Bravais lattice determination algorithm for rank $n > 3$.
	
	\item[2.] For $n = 2, 3$, the given algorithms are based on the Venkov reduction for $S_0 = A_n$ and $I_n$.
	Can a better algorithm be obtained by using the other reduction methods?
	
	\item[3.] Efficient algorithms for small $n$ to calculate the semi-metric and metric on ${\mathcal S}^n_{\succ 0} / GL_n(\ZZ)$ defined by Eq.(\ref{eq: semi-metric}) and Eq.(\ref{eq: distance}). Up to which $n$ can we enumerate all the necessary $g \in GL_n(\ZZ)$ to achieve the minimum value?

%For the former, it would be be possible to reduce the number of $g$ in the manner described in Section 6.1.
%For the latter, it would be possible to assume that both $S$ and $T$ are Minkowski-reduced and have the same determinant.

	\item[4.] Implement a metric on ${\mathcal S}^3_{\succ 0} / GL_3(\ZZ)$ that can be calculated only from the information in the theta series (\IE the values of $u S \tr{u}$, $u \in \ZZ^3$); Schiemann (1997)\nocite{Schiemann97} proved that 
	the following truncated theta series uniquely determines the class $[S] \in {\mathcal S}^3_{\succ 0} / GL_3(\ZZ)$ of $3$-by-$3$ metric tensor $S$:
	$$
	\Theta_S(z) = \sum_{u \in \ZZ^3,\ u S \tr{u} \le b(S)} e^{2 \pi i z (u S \tr{u})},
	$$
	where $b(S)$ is a constant calculated from the entries of Minkowski-reduced $S$ (Theorem 4.4).
	
	\item[5.] 
	Conway \& Sloane (1992) conjectured that the following map (vonorm map) uniquely determines the class $[S] \in {\mathcal S}^n_{\succ 0} / GL_n(\ZZ)$ of $n$-by-$n$ metric tensor $S$, and proved this for $n \le 3$:
	\begin{eqnarray*}
		{\rm vo}_S: \ZZ^n / 2 \ZZ^n & \rightarrow & \RR \\
		v + 2\ZZ^n  & \mapsto & \min\{ u S \tr{u} : u \in v + 2\ZZ^n \}.
	\end{eqnarray*}

	This conjecture has been proven to be true for 4D lattices \cite{Vallentin2003} and 5D lattices \cite{Sikiric2022}.
	
\end{itemize}

\section{Conclusion}
\label{Conclusion}

An error-stable Bravais lattice determination algorithm for 2D lattices was presented
along with theorems showing that it works under a mild condition on the magnitude of errors.
%The proof requires an assumption on the magnitude of errors in the input metric tensor, which was also used in our previous paper for 3D lattices. 
The entire algorithm for 3D lattices was also presented, which was implemented in the ab-initio indexing software \textit{CONOGRAPH} for powder and electron backscatter diffraction. 
Some methods to compare unit cells that have been used in crystallography and number theory were also introduced.

\paragraph{Acknowledgments}
The project was financially supported by JSPS KAKENHI (19K03628) and JST-FOREST Program (JPMJFR2132). 
We thank Mr.\ T.\ Tanaka of Nippon Steel Corporation for providing us with the experimental EBSD image.
The author would like to thank Prof. Kamiyama of High Energy Accelerator Research Organization for offering the CIF files of his Z-Database for testing codes.

% References are at the end of the document, between \begin{references}
	% and \end{references} tags. Each reference is in a \reference entry.

% \begin{references}
	% \reference{Author, A. \& Author, B. (1984). \emph{Journal} \textbf{Vol}, 
		% first page--last page.}
	% \end{references}
% \cite{knuth84}

%% Note added by Overleaf: If using bibtex, remove the "references" environment above, and uncomment the following line.
%\referencelist{iucr}

\bibliographystyle{apalike}
\bibliography{iucr}

\begin{thebibliography}{}

\bibitem[Andrews and Bernstein, 1988]{Andrews88}
Andrews, L.~C. and Bernstein, H.~J. (1988).
\newblock Lattices and reduced cells as points in 6-space and selection of
  bravais lattice type by projections.
\newblock {\em Acta Cryst.}, A44:1009--1018.

\bibitem[Andrews et~al., 2019]{Andrews2019a}
Andrews, L.~C., Bernstein, H.~J., and Sauter, N.~K. (2019).
\newblock Selling reduction versus niggli reduction for crystallographic
  lattices.
\newblock {\em Acta Crystallogr}, A75:115--120.

\bibitem[Andrews et~al., 2023]{Andrews2023}
Andrews, L.~C., Bernstein, H.~J., and Sauter, N.~K. (2023).
\newblock {SELLA}--a program for determining bravais lattice types
  (arxiv:2303.03122).

\bibitem[Anosova and Kurlin, 2021]{10.1007/978-3-030-76657-3_16}
Anosova, O. and Kurlin, V. (2021).
\newblock An isometry classification of periodic point sets.
\newblock In Lindblad, J., Malmberg, F., and Sladoje, N., editors, {\em
  Discrete Geometry and Mathematical Morphology}, pages 229--241, Cham.
  Springer International Publishing.

\bibitem[Aroyo, 2016]{Aroyo2016}
Aroyo, M.~I., editor (2016).
\newblock {\em International tables for crystallography}, volume~A.
\newblock Wiley.

\bibitem[Balashov and Ursell, 1957]{Balashov57}
Balashov, V. and Ursell, H.~D. (1957).
\newblock The choice of the standard unit cell in a triclinic lattice.
\newblock {\em Acta Cryst.}, 10:582--589.

\bibitem[Boultif and Lou{\"e}r, 2004]{Boultif2004}
Boultif, A. and Lou{\"e}r, D. (2004).
\newblock Powder pattern indexing with the dichotomy method.
\newblock {\em J. Appl. Cryst.}, 37:724--731.

\bibitem[Buerger, 1957]{Buerger57}
Buerger, M.~J. (1957).
\newblock Reduced cells.
\newblock {\em Z. Kristallogr.}, 109:42--60.

\bibitem[Burzlaff and Zimmermann, 1985]{Burzlaff85}
Burzlaff, H. and Zimmermann, H. (1985).
\newblock On the metrical properties of lattices.
\newblock {\em Z. Kristallogr.}, 170:247--262.

\bibitem[Cassels, 1978]{Cassels78}
Cassels, J. W.~S. (1978).
\newblock {\em Rational quadratic forms}.
\newblock Academic Press, London/New York.

\bibitem[Clegg, 1981]{Clegg81}
Clegg, W. (1981).
\newblock Cell reduction and lattice symmetry determination.
\newblock {\em Acta Cryst.}, A37:913--915.

\bibitem[Coelho, 2003]{Coelho2003}
Coelho, A.~A. (2003).
\newblock Indexing of powder diffraction patterns by iterative use of singular
  value decomposition.
\newblock {\em J. Appl. Cryst.}, 36:86--95.

\bibitem[Conway, 1997]{Conway97}
Conway, J.~H. (1997).
\newblock {\em The sensual (quadratic) form}.
\newblock Carus Mathematical Monographs 26, Mathematical Association of
  America.

\bibitem[Conway and Sloane, 1992]{Conway92}
Conway, J.~H. and Sloane, N. J.~A. (1992).
\newblock Low-dimensional lattices. vi. voronoi reduction of three-dimensional
  lattices.
\newblock {\em Proceedings: Mathematical and Physical Sciences}, 436:55--68.

\bibitem[de~Wolff, 1957]{Wolff57}
de~Wolff, P.~M. (1957).
\newblock On the determination of unit-cell dimensions from powder diffraction
  patterns.
\newblock {\em Acta Cryst.}, 10:590--595.

\bibitem[de~Wolff, 1958]{Wolff58}
de~Wolff, P.~M. (1958).
\newblock Detection of simultaneous zone relations among powder diffraction
  lines.
\newblock {\em Acta Cryst.}, 11:664--665.

\bibitem[de~Wolff, 1968]{Wolff68}
de~Wolff, P.~M. (1968).
\newblock A simplified criterion for the reliability of a powder pattern
  indexing.
\newblock {\em J. Appl. Cryst.}, 1:108--113.

\bibitem[Delaunay, 1933]{Delaunay33}
Delaunay, B. (1933).
\newblock Neue darstellung der geometrischen kristallographie.
\newblock {\em Z. Kristallogr.}, 84:109--149.

\bibitem[Eisenstein, 1851]{Eisenstein1851}
Eisenstein, G. (1851).
\newblock {T}abelle der reducirten positiven tern{\" a}ren quadratischen
  {F}ormen: nebst den {R}esultaten neuer {F}orschungen {\" u}ber diese formen.
\newblock {\em J. f. d. reine u. angew. Math.}, 41:140--190.

\bibitem[Fincke and Pohst, 1983]{Fincke83}
Fincke, U. and Pohst, M. (1983).
\newblock {\em On reduction algorithms in nonlinear integer mathematical
  programming in Operations research proceedings}.
\newblock Springer, Berlin.

\bibitem[Gilmore et~al., 2019]{Gilmore2019}
Gilmore, C.~J., Kaduk, J.~A., and Schenk, H., editors (2019).
\newblock {\em International tables for crystallography}, volume~H.
\newblock Wiley.

\bibitem[Grosse-Kunstleve et~al., 2004]{Kunstleve2004}
Grosse-Kunstleve, R.~W., Sauter, N.~K., and Adams, P.~D. (2004).
\newblock Numerically stable algorithms for the computation of reduced unit
  cells.
\newblock {\em Acta Cryst.}, A60:1--6.

\bibitem[Gruber, 1973]{Gruber73}
Gruber, B. (1973).
\newblock The relationship between reduced cells in a general bravais lattice.
\newblock {\em Acta Cryst.}, A29:433--440.

\bibitem[Gruber and Lekkerkerker, 1987]{Gruber87}
Gruber, P. and Lekkerkerker, C.~G. (1987).
\newblock {\em Geometry of Numbers}.
\newblock 2nd Edition, Elsevier.

\bibitem[Ito, 1949]{Ito49}
Ito, T. (1949).
\newblock A general powder x-ray photography.
\newblock {\em Nature}, 164:755--756.

\bibitem[Kohlbeck and Horl, 1978]{Kohlbeck78}
Kohlbeck, F. and Horl, E. (1978).
\newblock Trial and error indexing program for powder patternsof monoclinic
  substances.
\newblock {\em J. Appl. Cryst.}, 11:60--61.

\bibitem[K{\v r}iv{\' y} and Gruber, 1976]{Krivy76}
K{\v r}iv{\' y}, I. and Gruber, B. (1976).
\newblock A unified algorithm for determining the reduced (niggli) cell.
\newblock {\em Acta Cryst.}, A32:297--298.

\bibitem[Kurlin, 2022]{Kurlin2022}
Kurlin, V. (2022).
\newblock Mathematics of 2-dimensional lattices.
\newblock {\em Foundations of Computational Mathematics}.

\bibitem[Lang, 1998]{Lang98}
Lang, S. (1998).
\newblock {\em Fundamentals of Differential Geometry}, volume 191 of {\em
  Graduate Texts in Mathematics}.
\newblock Springer.

\bibitem[Lang, 1999]{Lang99}
Lang, S. (1999).
\newblock {\em Math Talks for Undergraduates}.
\newblock Springer.

\bibitem[Le~Bail, 2004]{LeBail2004}
Le~Bail, A. (2004).
\newblock Monte carlo indexing with {M}c{M}aille.
\newblock {\em Powder Diffraction}, 19:249--254.

\bibitem[Le~Page, 1982]{LePage82}
Le~Page, Y. (1982).
\newblock The derivation of the axis of the conventional unit cell from the
  dimensions of the buerger reduced cell.
\newblock {\em J. Appl. Cryst.}, 15:255--259.

\bibitem[Lenstra et~al., 1982]{Lenstra82}
Lenstra, A.~K., Lenstra~Jr., H.~W., and Lov{\' a}sz, L. (1982).
\newblock Factoring polynomials with rational coefficients.
\newblock {\em Mathematische Annalen}, 261(4):515--534.

\bibitem[Li and Han, 2015]{Li2015}
Li, L. and Han, M. (2015).
\newblock Determining the bravais lattice using a single electron backscatter
  diffraction pattern.
\newblock {\em J. Appl. Cryst.}, 48:107--115.

\bibitem[Li et~al., 2014]{Li2014}
Li, L., Ouyang, S., Yang, Y., and Han, M. (2014).
\newblock Ebsdl: a computer program for determining an unknown bravais lattice
  using a single electron backscatter diffraction pattern.
\newblock {\em J. Appl. Cryst.}, 47:1466--1468.

\bibitem[Michel, 1995]{Michel94}
Michel, L. (1995).
\newblock {\em Bravais classes, Vorono{\" i} cells, Delone symbols. in Symmetry
  and structural properties of condensed matter}.
\newblock Proceedings of the third international school on theoretical physics.
  World Scientific.

\bibitem[Minkowski, 1887]{Minkowski1887}
Minkowski, H. (1887).
\newblock Zur theorie der positiven quadratischen formen.
\newblock {\em J.Crelle}, 101:196--202.

\bibitem[Minkowski, 1905]{Minkowski05}
Minkowski, H. (1905).
\newblock Diskontinuit{\"a}tsbereich f{\"u}r arithmetische {\"a}quivalenz.
\newblock {\em J. Reine Angew. Math.}, 129:220--274.

\bibitem[Neumann, 2003]{Neumann2003}
Neumann, M.~A. (2003).
\newblock X-cell: a novel indexing algorithm for routine tasks and difficult
  cases.
\newblock {\em J. Appl. Cryst.}, 36(4):356--365.

\bibitem[Niggli, 1928]{Niggli28}
Niggli, P. (1928).
\newblock {\em Kristallographische und strukturtheoretische grundbegriffe.
  Handbuch der experimentalphysik}, volume~7.
\newblock Leipzig: Akademische Verlagsgesellschaft.

\bibitem[Nolze et~al., 2021]{Nolze2021}
Nolze, G., Tokarski, T., Rych\l{}owowski, L., Cios, G., and Winkelmann, A.
  (2021).
\newblock Crystallographic analysis of the lattice metric (calm) from single
  electron backscatter diffraction or transmission kikuchi diffraction
  patterns.
\newblock {\em J. Appl. Cryst.}, 54:1012--1022.

\bibitem[Oishi-Tomiyasu, 2012]{Tomiyasu2012}
Oishi-Tomiyasu, R. (2012).
\newblock Rapid bravais-lattice determination algorithm for lattice parameters
  containing large observation errors.
\newblock {\em Acta Cryst. A.}, 68:525--535.

\bibitem[Oishi-Tomiyasu, 2013]{Tomiyasu2013a}
Oishi-Tomiyasu, R. (2013).
\newblock Distribution rules of systematic absences on the conway topograph and
  their application to powder auto-indexing.
\newblock {\em Acta Cryst. A.}, 69:603--610.

\bibitem[Oishi-Tomiyasu, 2014]{Tomiyasu2014}
Oishi-Tomiyasu, R. (2014).
\newblock Robust powder auto-indexing using many peaks.
\newblock {\em J. Appl. Cryst.}, 47(2):593--598.

\bibitem[Oishi-Tomiyasu et~al., 2021]{Tomiyasu2021}
Oishi-Tomiyasu, R., Tanaka, T., and Nakagawa, J. (2021).
\newblock Distribution rules of systematic absences and generalized de wolff
  figures of merit applied to electron backscatter diffraction ab initio
  indexing.
\newblock {\em J. Appl. Cryst.}, 54:624--635.

\bibitem[Plesken and Souvignier, 1997]{Plesken97}
Plesken, W. and Souvignier, B. (1997).
\newblock Computing isometries of lattices.
\newblock {\em J. Symbolic Comp.,}, 24(3):327--334.

\bibitem[Quang and Murino, 2017]{Quang2017}
Quang, M.~H. and Murino, V. (2017).
\newblock {\em Covariances in Computer Vision and Machine Learning}.
\newblock Synthesis Lectures on Computer Vision. Morgan \& Claypool.

\bibitem[Ry{\u s}kov and Baranovski{\u i}, 1976]{Ryshkov76}
Ry{\u s}kov, S.~S. and Baranovski{\u i}, E.~P. (1976).
\newblock C-types of n-dimensional lattices and 5-dimensional primitive
  parallelohedra (with application to the theory of coverings).
\newblock {\em Proceedings of the Steklov Institute of Mathematics}, 137.

\bibitem[Schiemann, 1997]{Schiemann97}
Schiemann, A. (1997).
\newblock Ternary positive definite quadratic forms are determined by their
  theta series.
\newblock {\em Mathematische Annalen}, 308:507--517.

\bibitem[Selling, 1874]{Selling1874}
Selling, E. (1874).
\newblock {\"U}ber die bin{\"a}ren und tern{\"a}ren quadratischen formen.
\newblock {\em J. Reine Angew. Math.}, 77:143--229.

\bibitem[Sikiri{\' c} and Kummer, 2022]{Sikiric2022}
Sikiri{\' c}, M.~D. and Kummer, M. (2022).
\newblock Iso edge domains.
\newblock {\em Expositiones Mathematicae}, 40:302--314.

\bibitem[Terras, 1988]{Terras88}
Terras, A. (1988).
\newblock {\em Harmonic Analysis on Symmetric Spaces and Applications II}.
\newblock Springer Verlag.

\bibitem[Togashi, 2019]{Togashi2019}
Togashi, S. (2019).
\newblock Application of the theory of lattice-basis reduction to two lattice
  problems (in japanese).
\newblock Master's thesis, Yamagata University.

\bibitem[Vallentin, 2003]{Vallentin2003}
Vallentin, F. (2003).
\newblock {\em Sphere coverings, lattices, and tilings (in low dimensions)}.
\newblock PhD thesis, Technical University Munich, Germany.

\bibitem[Visser, 1969]{Visser69}
Visser, J.~W. (1969).
\newblock A fully automatic program for finding the unit cell from powder data.
\newblock {\em J. Appl. Cryst.}, 2:89--95.

\bibitem[Voronoi, 1908]{Voronoi07}
Voronoi, G.~F. (1908).
\newblock Sur quelques proprietes des formes quadratiques positives parfaites.
\newblock {\em J. Reine Angew. Math.}, 133:97--178.

\bibitem[Widdowson and Kurlin, 2022]{NEURIPS2022_9c256fa1}
Widdowson, D. and Kurlin, V. (2022).
\newblock Resolving the data ambiguity for periodic crystals.
\newblock In Koyejo, S., Mohamed, S., Agarwal, A., Belgrave, D., Cho, K., and
  Oh, A., editors, {\em Advances in Neural Information Processing Systems},
  volume~35, pages 24625--24638. Curran Associates, Inc.

\bibitem[Zhilinskii, 2016]{Zhilinskii2016}
Zhilinskii, B. (2016).
\newblock {\em Introduction to Louis Michel's Lattice Geometry Through Group
  Action (Current Natural Sciences)}.
\newblock EDP Sciences.

\bibitem[Zimmermann and Burzlaff, 1985]{Zimmermann85}
Zimmermann, H. and Burzlaff, H. (1985).
\newblock Delos--a computer program for the determination of a unique
  conventional cell.
\newblock {\em Z. Kristallogr.}, 170:241--246.

\bibitem[Zuo et~al., 1995]{Zuo95}
Zuo, L., Muller, J., Philippe, M.~J., and Esling, C. (1995).
\newblock A refined algorithm for the reduced-cell determination.
\newblock {\em Acta Cryst.}, A51:943--945.

\end{thebibliography}

%-------------------------------------------------------------------------
% TABLES AND FIGURES SHOULD BE INSERTED AFTER THE MAIN BODY OF THE TEXT
%-------------------------------------------------------------------------

% Simple tables should use the tabular environment according to this
% model

\end{document}